\documentclass[aps,preprint,nofootinbib]{revtex4}
\usepackage{amsmath}
\usepackage{graphicx}
\usepackage{color}
\usepackage{slashed}
\usepackage{subfigure}
\usepackage{tabularx}
\usepackage{booktabs}

\allowdisplaybreaks

\begin{document}

\title{$B_c$ meson exclusive decays to a $P$-wave charmonium and a pion at NLO accuracy\\[0.7cm]}

\author{\vspace{1cm} Zi-Qiang Chen$^{1,2}$\footnote[2]{
chenziqiang13@mails.ucas.ac.cn} and Cong-Feng Qiao$^{1,2}$\footnote[1]{qiaocf@ucas.ac.cn, corresponding author} \\}

\affiliation{$^1$ School of Physics, University of Chinese Academy of
Sciences, Beijing 100049, China\\
$^2$ CAS Center for Excellence in Particle Physics, Beijing 10049, China
\vspace{0.6cm}}

\begin{abstract}
\vspace{0.5cm}
In this paper, we calculate the next-to-leading order (NLO) quantum chromodynamics (QCD) corrections to the exclusive processes $B_c^+\to \chi_{cJ}(h_c)\pi^+$ in the framework of the nonrelativistic QCD (NRQCD) factorization formalism.
The results show that NLO QCD corrections markedly enhance the branching ratios with $K$ factors of about 2.5.
In combination with the study of $B_c^+\to J/\psi\pi^+$, we find that the NLO NRQCD prediction for the ratio of branching fractions $\frac{\mathcal{B}(B_c^+\to \chi_{c0}\pi^+)}{\mathcal{B}(B_c^+\to J/\psi \pi^+)}$ is then compatible with the experimental measurement.
\end{abstract}
\maketitle

\newpage

\section{Introduction}
In the Standard Model (SM) of particle physics, the $B_c$ meson family is unique as its states are composed of two different heavy quarks, beauty ($b$) and charm ($c$).
Unlike the charmonium and bottomonium states, the ground state of $B_c$ meson family decays only via weak interactions,
where the main decay modes can be classified as: i) $\bar{b}$ decay with $c$ as a spectator, ii) $c$ decay with $\bar{b}$ as a spectator, and iii) $\bar{b}c$ annihilate to a virtual $W$ boson.
Study on these decay processes can deepen our understanding of both strong and weak interactions, and provides opportunities to search for new physics beyond the SM.

Experimentally, the $B_c$ meson decay to charmonium processes are essential for the reconstruction of $B_c$ meson signals.
After the discovery of ground state $B_c$ by the CDF Collaboration in 1998 \cite{Abe:1998wi},
there have been continuous measurements on its lifetime \cite{Abulencia:2006zu,Abazov:2008rba} and mass \cite{Aaltonen:2007gv,Abazov:2008kv,Aaij:2012dd,Aaij:2020jrx} via the semileptonic decay $B_c^+\to J/\psi l^+\nu_l$ and the exclusive two-body decay $B_c^+\to J/\psi\pi^+$.
In 2014, the ATLAS Collaboration reported an excited state with mass of $6842\pm4\pm5$ MeV \cite{Aad:2014laa}, which is regarded as a candidate of $B_c(2S)$ state. This observation was confirmed by the CMS and LHCb Collaborations in 2019, where the $B_c^+(2S)$ and $B_c^{*+}(2S)$ states were reconstructed through the decays $B_c^+(2S)\to B_c^+\pi^+\pi^-$ and $B_c^{*+}(2S)\to B_c^{*+}(\to B_c^+\gamma)\pi^+\pi^-$ \cite{Sirunyan:2019osb,Aaij:2019ldo}, respectively.
Note, in these works, the intermediate $B_c^+$ was also reconstructed through $B_c^+\to J/\psi\pi^+$.

Besides $B_c^+$ decay to $S$-wave charmonium, its transition to $P$-wave charmonium is also interesting for three reasons:
\begin{enumerate}
\item The branching ratio $\mathcal{B}(\chi_{c2}\to J/\psi\gamma)=19.0\pm0.5\%$ \cite{Zyla:2020zbs} is sizable\footnote{Although $\mathcal{B}(\chi_{c1}\to J/\psi\gamma)=34.3\pm 1.0\%$ is also sizable, theoretical prediction indicates that $\mathcal{B}(B_c^+\to \chi_{c1}\pi^+)$ is insignificant compared to $\mathcal{B}(B_c^+\to J/\psi\pi^+)$.}.
Hence the cascade decay $B_c^+\to \chi_{c2}(\to J/\psi\gamma)\pi^+$ may contribute a substantial background for the $B_c\to J/\psi\pi^+$ process.
\item The $\bar{b}c$ annihilation decay of $B_c^+$ is an interesting topic in $B$ physics.
Any observation of significant enhancement over the SM prediction could indicate the presence of new physics effects.
The decay of $B_c^+$ to three light charged hardrons, like $B_c^+\to K^+K^-\pi^+$, provides a good way to study this issue.
The contributions from intermediate states, like $B_c^+\to \chi_{c0}(\to K^+K^-)\pi^+$ should be subtracted to obtain the annihilation contribution \cite{Aaij:2016xas}.
\item Although the exact values of the branching ratios for $B_c^+\to J/\psi \pi^+$ and $B_c^+\to \chi_{cJ}(h_c)\pi^+$ have not been measured yet,
the ratio $\frac{\mathcal{B}(B_c^+\to \chi_{c0}\pi^+)}{\mathcal{B}(B_c^+\to J/\psi \pi^+)}$ is accessible by combining the results of Refs. \cite{Zyla:2020zbs,Aaij:2016xas,Aaij:2014ija}.
Phenomenological study on this issue provides an opportunity to test the nonrelativistic QCD (NRQCD) effective theory.
\end{enumerate}

Theoretically, the exclusive two-body decay of $B_c^+$ into $P$-wave charmonium and a light meson has been studied in various approaches:
the NRQCD approach \cite{Kiselev:2001zb,Zhu:2017lwi}, the perturbative QCD approach \cite{Rui:2017pre}, the Bethe-Salpeter equation \cite{Chang:2001pm,Wang:2011jt}, the nonrelativistic quark model \cite{Hernandez:2006gt}, and the relativistic quark model \cite{Ivanov:2006ni,Ebert:2010zu}.
We notice that their predictions on the branching ratios are generally incompatible with each other, and their analyses are limited to the leading-order (LO) accuracy.
Considering the fact that the higher-order QCD corrections in quarkonium energy regime are normally significant \cite{Qiao:2012hp}, in this work we calculate the next-to-leading order (NLO) QCD corrections to $B_c^+\to \chi_{cJ}(h_c)\pi^+$ processes in the framework of NRQCD factorization formalism \cite{Bodwin:1994jh}.

The rest of the paper is organized as follows.
In Sec. II, we present the primary formulas employed in the calculation.
In Sec. III, we elucidate some technical details for the analytical calculation.
In Sec. IV, the numerical evaluation for concerned processes is performed at NLO QCD accuracy.
The last section is reserved for summary and conclusions.

\section{Formalism}
\subsection{Effective weak Hamiltonian}
In the SM, $B^+_c\to \chi_{cJ}(h_c)\pi^+$ occur through $W$-mediated charge-current processes.
However, since $m_W\gg m_{b,c}$, large logarithm terms will arise in higher-order QCD corrections.
Thus, the renormalization-group (RG) improved effective weak Hamiltonian method is usually employed in the calculation.
The interaction term is
\begin{equation}
\mathcal{H}_{\rm eff}=\frac{G_F}{\sqrt{2}}V_{ud}V^*_{cb}\left(C_1(\mu)\mathcal{Q}_1(\mu)+C_2(\mu)\mathcal{Q}_2(\mu)\right),
\end{equation}
where $G_F$ is the Fermi constant, $V_{ud}$ and $V_{cb}$ are the Cabibbo-Kobayashi-Maskawa (CKM) matrix elements; $C_{1,2}(\mu)$ are the perturbatively calculable Wilson coefficients, $\mathcal{Q}_{1,2}(\mu)$ are the local four-quark operators, which take the form
\begin{align}
&\mathcal{Q}_1(\mu)=\bar{b}_i\gamma^\mu(1-\gamma_5)c_i\otimes \bar{u}_j\gamma_\mu(1-\gamma_5)d_j,\\
&\mathcal{Q}_2(\mu)=\bar{b}_i\gamma^\mu(1-\gamma_5)c_j\otimes \bar{u}_j\gamma_\mu(1-\gamma_5)d_i.
\end{align}
Here $i$, $j$ are color indices and the summation convention for repeated indices is understood.
In contrast to conventional operators $\mathcal{Q}_{1,2}$, we will adopt another basis
\begin{align}
&\mathcal{Q}_0(\mu)=\bar{b}_i\gamma^\mu(1-\gamma_5)c_i\otimes \bar{u}_j\gamma_\mu(1-\gamma_5)d_j, \\
&\mathcal{Q}_8(\mu)=\bar{b}_i T^a_{ij}\gamma^\mu(1-\gamma_5)c_j\otimes \bar{u}_k T^a_{kl}\gamma_\mu(1-\gamma_5)d_l,
\label{eq_4quarkop}%
\end{align}
where $T^a$ is the generator of $SU(3)$ fundamental representation.
By applying the Fierz rearrangement relation
\begin{equation}
T^A_{ij}T^A_{kl}=-\frac{1}{6}\delta_{ij}\delta_{kl}+\frac{1}{2}\delta_{il}\delta_{kj},
\end{equation}
we obtain
\begin{equation}
\mathcal{Q}_0=\mathcal{Q}_1,\quad \mathcal{Q}_8=-\frac{1}{6}\mathcal{Q}_1+\frac{1}{2}\mathcal{Q}_2,
\end{equation}
and
\begin{equation}
C_0=C_1+\frac{1}{3}C_2,\quad C_8=2C_2.
\end{equation}
The Wilson coefficients $C_{0,8}$ can be obtained by solving the RG equation.
Under the leading logarithmic approximation,
\begin{align}
&C_0(\mu)=\frac{2}{3}\bigg[\frac{\alpha_s(m_W)}{\alpha_s(\mu)}\bigg]^{\frac{\gamma_+}{2\beta_0}}+\frac{1}{3}\bigg[\frac{\alpha_s(m_W)}{\alpha_s(\mu)}\bigg]^{\frac{\gamma_-}{2\beta_0}},\\
&C_8(\mu)=\bigg[\frac{\alpha_s(m_W)}{\alpha_s(\mu)}\bigg]^{\frac{\gamma_+}{2\beta_0}}-\bigg[\frac{\alpha_s(m_W)}{\alpha_s(\mu)}\bigg]^{\frac{\gamma_-}{2\beta_0}},
\end{align}
with
\begin{equation}
\gamma_\pm=\pm6\frac{N_c\mp 1}{N_c}.
\end{equation}
Here, $\alpha_s$ is the running coupling constant of QCD; $\beta_0=(11/3)C_A-(4/3)T_fn_f$ is the one-loop coefficient of QCD beta function, $n_f$ is the number of active quarks, and $C_A=N_c=3$, $T_f=1/2$ are normal color factors.

\subsection{Projection operators and decay amplitudes}
According to the NRQCD factorization formalism \cite{Bodwin:1994jh} and the factorization formalism for nonleptonic $B$ meson decay \cite{Beneke:2000ry}, the decay amplitude of $B_c^+\to \chi_{cJ}(h_c)\pi^+$ is conjectured to be factorized as
\begin{equation}
\mathcal{M}\left(B_c^+\to \chi_c(h_c)\pi^+\right)\simeq \frac{if_\pi}{8\pi}\frac{|R_{cc}^\prime(0)|}{\sqrt{8m_c^3}}\frac{|R_{bc}(0)|}{\sqrt{m_b+m_c}}\int_0^1 dx T(x,\mu)\phi_\pi(x,\mu),
\label{eq_factorization}
\end{equation}
where, $f_\pi$ is the decay constant of pion;
$R_{bc}(0)$ is the radial wave function at the origin for $B_c^+$, $R_{cc}^\prime(0)$ is the derivative of radial wave function for $\chi_{cJ}(h_c)$,
$T(x,\mu)$ is the perturbatively calculable hard kernel,
and $\phi_\pi(x,\mu)$ is the leading-twist light-cone distribution amplitude (LCDA) of pion.

With the decay amplitude $\mathcal{M}$, the branching ratio can be obtained through
\begin{equation}
\mathcal{B}=\frac{\tau_{B_c}^{}}{16\pi}\frac{(m_b+m_c)^2-4m_c^2}{(m_b+m_c)^3}|\mathcal{M}|^2,
\label{eq_Mtobr}
\end{equation}
where $\tau^{}_{B_c}$ is the mean life of $B_c^+$ meson.
At the NLO QCD accuracy, $|\mathcal{M}|^2\simeq|\mathcal{M}_{\rm born}|^2+2{\rm Re(\mathcal{M}_{\rm loop}\mathcal{M}_{\rm born}^*)}$, where $\mathcal{M}_{\rm born}$ is the Born amplitude at $\mathcal{O}(\alpha_s)$, and $\mathcal{M}_{\rm loop}$ is the one-loop correction at $\mathcal{O}(\alpha_s^2)$.

The hard kernel $T(x,\mu)$ can be computed by using the covariant projection operator method.
The spin and color projection operators used in our calculation are
\begin{align}
&\Pi_{bc}^0=\frac{1}{2}(\slashed{k}_1+m_b+m_c)\gamma_5\otimes \frac{1_c}{\sqrt{N_c}},\\
&\Pi_{cc}^{\mu}=\Big(\frac{\slashed{k}_2}{2}-\slashed{q}-m_c\Big)\gamma^\mu\Big(\frac{\slashed{k}_2}{2}+\slashed{q}+m_c\Big)\otimes \frac{1_c}{\sqrt{N_c}}, \\
&\Pi_{cc}^0=\Big(\frac{\slashed{k}_2}{2}-\slashed{q}-m_c\Big)\gamma_5\Big(\frac{\slashed{k}_2}{2}+\slashed{q}+m_c\Big)\otimes \frac{1_c}{\sqrt{N_c}}, \\
&\Pi_{ud}^0=\frac{1}{2}\slashed{k}_3\gamma_5\otimes \frac{1_c}{\sqrt{N_c}},
\end{align}
where $k_1$ denotes the momentum of $B_c^+$, $k_2$ the momentum of $\chi_{cJ}(h_c)$, $k_3$ the momentum of $\pi^+$, and $2q$ is the relative momentum between the $c\bar{c}$ pair.
Then the hard kernels can be expressed as
\begin{align}
&T(\chi_{c0})=\frac{1}{\sqrt{3}}I_{\alpha\beta}\frac{d}{dq_\beta}{\rm Tr}\big(\Pi_{cc}^\alpha \mathcal{A}\big)\bigg|_{q\to 0},\label{eq_proamp_a} \\
&T(\chi_{c1}(\lambda))=\frac{i}{2\sqrt{2}m_c}\varepsilon_{\alpha\beta\mu\nu}k_1^\mu\epsilon^\nu(\lambda)\frac{d}{dq_\beta}{\rm Tr}\big(\Pi_{cc}^\alpha \mathcal{A}\big)\bigg|_{q\to 0},\label{eq_proamp_b}\\
&T(\chi_{c2}(\lambda))=\bigg[\frac{1}{2}(I_{\alpha\mu}I_{\beta\nu}+I_{\alpha\nu}I_{\beta\mu})-\frac{1}{3}I_{\alpha\beta}I_{\mu\nu}\bigg]\epsilon^{\mu\nu}(\lambda)\frac{d}{dq_\beta}{\rm Tr}\big(\Pi_{cc}^\alpha \mathcal{A}\big)\bigg|_{q\to 0}, \label{eq_proamp_c} \\
&T(h_c(\lambda))=\epsilon_\beta(\lambda)\frac{d}{dq_\beta}{\rm Tr}\big(\Pi_{cc}^0 \mathcal{A}\big)\bigg|_{q\to 0}.
\label{eq_proamp_d}%
\end{align}
Here, $T(H(\lambda))$ denotes the hard kernel referring to $B_c^+\to H(\lambda)\pi^+$ process, with $\lambda$ labeling the helicity state of $H$, $\epsilon(\lambda)$ labeling the corresponding polarization vector;
$\mathcal{A}$ denotes the standard amplitude for partonic process, amputated of quark spinors.
Note, in Eqs. \eqref{eq_proamp_a}$\sim$\eqref{eq_proamp_d}, the projectors $\Pi_{bc}^0$ and $\Pi_{ud}^0$ are suppressed for brevity.
The tensor $I_{\alpha\beta}$ in Eqs. \eqref{eq_proamp_a} and \eqref{eq_proamp_c} is
\begin{equation}
I_{\alpha\beta}=-g_{\alpha\beta}+\frac{k_{2\alpha}k_{2\beta}}{4m_c^2}.
\end{equation}

Up to NLO, the hard kernel $T(x,\mu)$ can be written as
\begin{align}
T(x,\mu)=\frac{G_F}{\sqrt{2}}&V_{ud}V^*_{cb}\frac{\alpha_s(\mu)}{\pi}\bigg\{C_0(\mu)T_0^{(0)}(x)+C_8(\mu)T_8^{(0)}(x)+\frac{\alpha_s(\mu)}{\pi}\bigg[C_0(\mu)T_0^{(1)}(x)\nonumber \\
&+C_8(\mu)T_8^{(1)}(x)+\Big(C_0(\mu)\tilde{T}_0^{(1)}(x)+C_8(\mu)\tilde{T}_8^{(1)}(x)\Big){\rm ln}\frac{\mu^2}{(m_b+m_c)^2}\bigg]\bigg\},
\label{eq_Tpertur}
\end{align}
where $T^{(j)}_i$ and $\tilde{T}^{(j)}_i$ are dimensionless and can only depend on $x$ and $m_c/m_b$.
Note, here and throughout, the renormalization scale is set equal to the factorization scale.

\section{Analytical calculation}
\label{sec_ana}
\subsection{Kinematics and LO calculation}

\begin{figure}
\includegraphics[width=\textwidth]{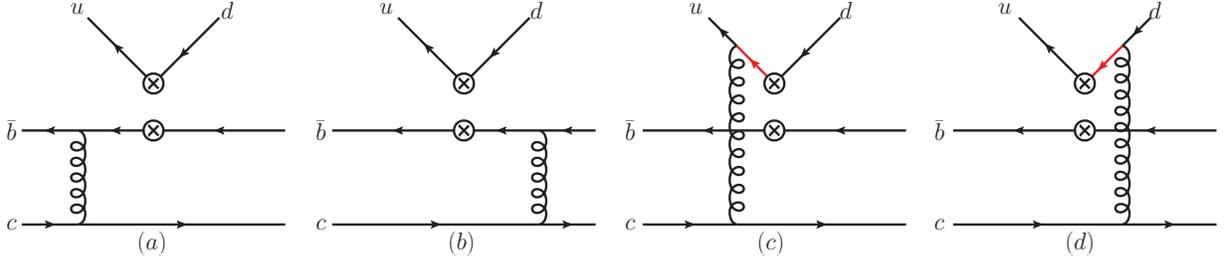}
\caption{Tree-level Feynman diagrams for $B_c^+\to \chi_c(h_c)\pi^+$. The vertex ``$\otimes\otimes$'' denotes the insertion of effective weak interaction.}
\label{fig_Feytree}
\end{figure}

The tree-level Feynman diagrams for partonic processes are shown in Fig. \ref{fig_Feytree}.
Therein, (a) and (b) contribute to $T^{(0)}_0$, (c) and (d) contribute to $T^{(0)}_8$.
The momenta of incoming and outgoing particles are denoted as
\begin{equation}
[\bar{b}c](k_1)\to [c\bar{c}](k_2,\epsilon)+[u\bar{d}](k_3).
\end{equation}
Here, the initial and final state particles are all on their mass shells: $k_1^2=(m_b+m_c)^2$, $k_2^2=4m_c^2$, $k_3^2=0$.
By introducing the orthonormal four-vector base: $n_0=(1,0,0,0)$, $n_1=(0,1,0,0)$, $n_2=(0,0,1,0)$, $n_3=(0,0,0,1)$, the momenta of external particles in the initial state rest frame can be assigned as
\begin{align}
&k_1^\mu=(m_b+m_c)n_0^\mu, \nonumber \\
&k_2^\mu=\frac{m_b+m_c}{2}[(4r^2-8r+5)n_0^\mu-(3-2r)(2r-1)n_3^\mu],\nonumber\\
&k_3^\mu=\frac{m_b+m_c}{2}(3-2r)(2r-1)(n_0^\mu+n_3^\mu).
\end{align}
Here we introduce $r=\frac{m_b}{m_b+m_c}$.
For convenience, we also introduce
\begin{align}
&e_1^\mu=n_1^\mu,\quad e_2^\mu=n_2^\mu,\nonumber \\
&e_3^\mu=\frac{1}{4(1-r)}[(3-2r)(2r-1)n_0^\mu+(4r^2-8r+5)n_3^\mu].
\end{align}
Then the polarization vectors for $\chi_{c1}$ and $h_c$ can be chosen as
\begin{equation}
\epsilon^\mu(1)=e_1^\mu,\quad \epsilon^\mu(2)=e_2^\mu,\quad \epsilon^\mu(3)=e_3^\mu.
\end{equation}
The polarization vectors for $\chi_{c2}$ can be constructed as
\begin{align}
&\epsilon^{\mu\nu}(1)=\frac{1}{\sqrt{2}}(e_1^\mu e_2^\nu+e_2^\mu e_1^\nu),\quad \epsilon^{\mu\nu}(2)=\frac{1}{\sqrt{2}}(e_1^\mu e_1^\nu-e_2^\mu e_2^\nu),\nonumber \\
&\epsilon^{\mu\nu}(3)=\frac{1}{\sqrt{2}}(e_1^\mu e_3^\nu+e_3^\mu e_1^\nu),\quad \epsilon^{\mu\nu}(4)=\frac{1}{\sqrt{2}}(e_2^\mu e_3^\nu+e_3^\mu e_2^\nu),\nonumber \\
&\epsilon^{\mu\nu}(5)=\frac{1}{\sqrt{6}}(e_1^\mu e_1^\nu+e_2^\mu e_2^\nu-2e_3^\mu e_3^\nu).
\end{align}
In fact,  according to the helicity conservation rule \cite{Brodsky:1981kj}, the only nonzero helicity amplitudes for $\chi_{c1}$, $\chi_{c2}$, and $h_c$ channels are $T(\chi_{c1}(3))$, $T(\chi_{c2}(5))$, and $T(h_c(3))$ respectively.

The tree-level calculation is straightforward, and the results are pretty simple:
\begin{align}
&T^{(0)}_0(\chi_{c0})=-\frac{256 \pi ^2 (2 r-3) (4 r^2-4 r+3)}{(r-1) (2 r-1)^4},\label{eq_Ttree1}\\
&T^{(0)}_8(\chi_{c0})=\frac{64\pi^2}{9(2r-1)^2}\bigg[\frac{8 r^2-14 r+7}{(2 r-1) (r-1)(x-x_1+i\varepsilon)}+\frac{4 (2 r^2+3 r-3)}{(2 r-1)(x-x_2-i\varepsilon)}\nonumber \\
&\quad\quad\quad\quad+\frac{2 (r-1)}{(2 r-3)(x-x_1+i\varepsilon)^2}+\frac{4 (r-1)^2}{(2 r-3)(x-x_2-i\varepsilon)^2}\bigg]; \label{eq_Ttree2}\\
&T^{(0)}_0(\chi_{c1})=-\frac{256 \sqrt{6} \pi ^2 (2 r-3)}{3(r-1) (2 r-1)^2},\label{eq_Ttree3}\\
&T^{(0)}_8(\chi_{c1})=\frac{64 \sqrt{6} \pi ^2 }{9 (2 r-1)^3}\bigg[\frac{4 r-3}{(r-1) (x-x_1+i\varepsilon)}-\frac{2 (4 r^2-6 r+3)}{x-x_2-i\varepsilon}\bigg];\label{eq_Ttree4}\\
&T^{(0)}_0(\chi_{c2})=-\frac{256 \sqrt{2} \pi ^2 (2 r-3)^2}{(r-1) (2 r-1)^4},\label{eq_Ttree5}\\
&T^{(0)}_8(\chi_{c2})=\frac{128\sqrt{2}\pi^2}{9(2r-1)^2}\bigg[\frac{4 r^2-10 r+5}{2(2 r-1) (r-1)(x-x_1+i\varepsilon)}-\frac{2 (2 r^2-6 r+3)}{(2 r-1)(x-x_2-i\varepsilon)}\nonumber \\
&\quad\quad\quad\quad-\frac{r-1}{(2 r-3)(x-x_1+i\varepsilon)^2}-\frac{2 (r-1)^2}{(2 r-3)(x-x_2-i\varepsilon)^2}\bigg];\label{eq_Ttree6}\\
&T^{(0)}_0(h_c)=-\frac{512 \sqrt{3}\pi ^2 (2 r-3) (4 r^3-8 r^2+7 r-5)}{(r-1) (2 r-1)^4},\label{eq_Ttree7}\\
&T^{(0)}_8(h_c)=\frac{128\sqrt{3}\pi^2}{9(2r-1)^2}\bigg[-\frac{1}{2(r-1)(x-x_1+i\varepsilon)}+\frac{2 (r-1)}{x-x_2-i\varepsilon}\nonumber \\
&\quad\quad\quad\quad-\frac{r-1}{(2 r-3)(x-x_1+i\varepsilon)^2}+\frac{2 (r-1)^2}{(2 r-3)(x-x_2-i\varepsilon)^2}\bigg].
\label{eq_Ttree8}%
\end{align}
Here, $x$ is the momentum fraction assigned to the $u$ quark, $x_1$ and $x_2$ are defined as
\begin{equation}
x_1=\frac{1-r}{3-2r},\quad\quad x_2=\frac{2-r}{3-2r}.
\end{equation}
For $m_b,m_c>0$, we have $0<x_1,x_2<1$.

The $\frac{1}{x-x_1+i\varepsilon}$ and $\frac{1}{x-x_2-i\varepsilon}$ terms in Eqs. \eqref{eq_Ttree2}, \eqref{eq_Ttree4}, \eqref{eq_Ttree6}, and \eqref{eq_Ttree8} arise from the light quark propagators of Fig. \ref{fig_Feytree}(c) and (d) (the red lines).
The $\frac{1}{(x-x_1+i\varepsilon)^2}$ and $\frac{1}{(x+x_2-i\varepsilon)^2}$ terms arise due to the derivative relative to $q$, and hence will absent in $B_c$ decay to $S$-wave charmonium process.
According to Eq. \eqref{eq_factorization}, the hard kernel $T(x,\mu)$ needs to be convoluted with LCDA $\phi_\pi(x,\mu)$ to obtain the decay  amplitude.
The integral interval $x\in [0,1]$ include the poles $x_1$ and $x_2$, while the singularities are tamed by the $i\varepsilon$ prescription
\begin{align}
&\int_0^1dx\frac{1}{x-x_i\pm i\varepsilon}=\ln\frac{1-x_i}{x_i}\mp i\pi,\quad {\rm for}\ 0<x_i<1,\\
&\int_0^1dx\frac{1}{(x-x_i\pm i\varepsilon)^2}=-\frac{1}{x_i(1-x_i)},\quad {\rm for}\ 0<x_i<1.
\end{align}
The above equations can be obtained by deforming the integral paths away from poles.
However, at the heavy quark limit $r\to 1$, the poles approach to the end points.
Considering the $x(1-x)$ factor in $\phi_\pi(x,\mu)$, we have
\begin{equation}
\int_0^1dx T(x,\mu)\phi_\pi(x,\mu)\overset{r\to 1}{\sim} A\int_0^1 dx\frac{1}{x}+B\int_0^1 dx\frac{1}{1-x},
\end{equation}
which is divergent and indicates a potential violation of the factorization formalism at the tree-level.

\subsection{NLO corrections}

\begin{figure}
\includegraphics[width=\textwidth]{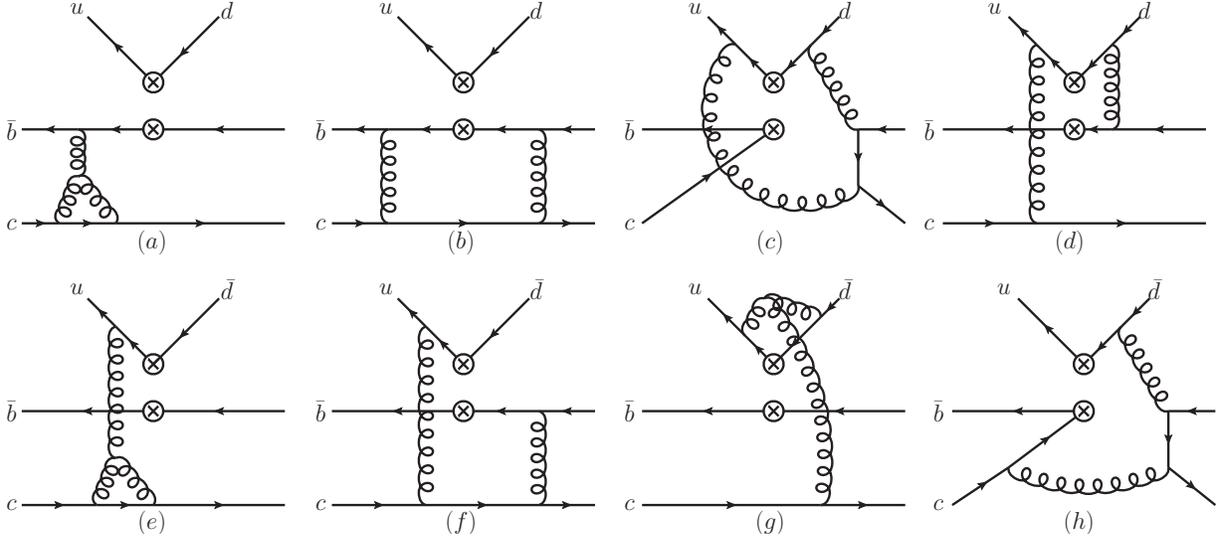}
\caption{Typical one-loop Feynman diagrams for $B_c^+\to \chi_c(h_c)\pi^+$. The vertex ``$\otimes\otimes$'' denotes the insertion of effective weak interaction.}
\label{fig_Feynlo}
\end{figure}

The typical one-loop Feynman diagrams are shown in Fig. \ref{fig_Feynlo}.
Therein, (a)$\sim$(c) contribute to $T^{(1)}_0$; (e)$\sim$(h) contribute to $T^{(1)}_8$; (d) contributes to both $T^{(1)}_0$ and $T^{(1)}_8$.
Note, according to the Ward-Takahashi identity \cite{Ward:1950xp,Takahashi:1957xn}, the contributions of type (c) diagrams are canceled by each other, which is verified by our explicit calculation.

In the computation of one-loop amplitudes, dimensional regularization with $D=4-2\epsilon$ is employed to regularize the ultraviolet (UV) and infrared (IR) singularities.
Explicitly, in our calculation, the momenta and polarization vectors of external particles are kept in four-dimensions, while other quantities are continued to $D$-dimensions.
The method proposed in Refs. \cite{Kreimer:1989ke,Korner:1991sx} is used to deal with the $D$-dimensional $\gamma_5$ trace.

The UV singularities are removed by the renormalization procedure.
The standard renormalization constants include $Z_2$, $Z_m$, $Z_l$, and $Z_g$, corresponding to heavy quark field, heavy quark mass, light quark field, and strong coupling constant, respectively.
We define $Z_2$, $Z_m$ and $Z_l$ in the on-shell (OS) scheme, $Z_g$ in the modified minimal-subtraction ($\overline{\rm MS}$) scheme.
The corresponding counterterms are
\begin{align}
&\delta Z_2^{\rm OS}=-C_F\frac{\alpha_s}{4\pi}\left[\frac{1}{\epsilon_{\rm UV}}+\frac{2}{\epsilon_{\rm IR}}-3\gamma_E+3\ln\frac{4\pi\mu^2}{m^2}+4\right],\\
&\delta Z_m^{\rm OS}=-3C_F\frac{\alpha_s}{4\pi}\left[\frac{1}{\epsilon_{\rm UV}}-\gamma_E+\ln\frac{4\pi\mu^2}{m^2} +\frac{4}{3}\right], \\
&\delta Z_l^{\rm OS}=-C_F\frac{\alpha_s}{4\pi}\left[\frac{1}{\epsilon_{\rm UV}}-\frac{1}{\epsilon_{\rm IR}}\right],\\
&\delta Z_g^{\overline{\rm MS}}=-\frac{\beta_0}{2}\frac{\alpha_s}{4\pi}\left[\frac{1}{\epsilon_{\rm UV}} -\gamma_E + \ln(4\pi)\right].
\end{align}
Here, $\gamma_E$ is the Euler's constant; $m$ stands for $m_c$ and $m_b$ accordingly.
Besides the standard QCD renormalization, the renormalization of effective weak Hamiltonian leads to the counterterm
\begin{equation}
\delta T_i^{(1)}=\delta Z^{\mathcal{Q}}_{ij}T^{(0)}_j,
\end{equation}
where $i,j=0,8$. At the $\overline{\rm MS}$ scheme \cite{Gambino:2003zm},
\begin{equation}
\delta Z^{\mathcal{Q}}_{ij}=\frac{1}{4}
\bigg(
  \begin{array}{cc}
    0 & 6\\
    4/3 &  -2\\
  \end{array}
\bigg)
\bigg[\frac{1}{\epsilon_{\rm UV}}-\gamma_E+\ln(4\pi)\bigg].
\end{equation}
Note, the overall constant factor of $\delta T_i^{(1)}$ is defined as the same as that of $T_i^{(1)}$ (see Eq. \eqref{eq_Tpertur}).
Hence comparing to the traditional form, the $\delta Z^{\mathcal{Q}}_{ij}$ here is divided by an additional factor, $\alpha_s/\pi$.

While for the IR singularities, parts of them are canceled each other, the remaining are canceled by the counterterm arising from the corrections of LCDA.
At the $\overline{\rm MS}$ scheme \cite{Braaten:1982yp},
\begin{align}
\delta T^{(1)}_i=\frac{C_F}{2}\bigg(\frac{1}{\epsilon_{\rm IR}}-\gamma_E&+\ln(4\pi)\bigg)\int _0^1 dy \big(T^{(0)}_i(y)-T^{(0)}_i(x)\big)\nonumber \\
&\bigg[\frac{y}{x}\bigg(1+\frac{1}{x-y}\bigg)\theta(x-y)+\frac{1-y}{1-x}\bigg(1+\frac{1}{y-x}\bigg)\theta(y-x)\bigg].
\end{align}
Since $T^{(0)}_0$ is independent of $x$, the counterterm to $T^{(1)}_0$ vanishes.

At the regions near $x_1$ and $x_2$, the asymptotic behavior of $T_i^{(1)}$ is $T_i^{(1)}\sim \ln^m(x-x_1+i\varepsilon)/(x-x_1+i\varepsilon)^n$ and $T_i^{(1)}\sim \ln^m(x-x_2-i\varepsilon)/(x-x_2-i\varepsilon)^n$ respectively.
Since $x_1$ and $x_2$ are not pinched, the convolution of Eq. \eqref{eq_factorization} is free from singularities.

\section{Numerical results}
\subsection{Parameters and decay amplitudes}
The input parameters taken in the numerical calculation go as follows \cite{Zyla:2020zbs}:
\begin{align}
&G_F=1.166\times 10^{-5}\ {\rm GeV}^{-2},\quad |V_{ud}|=0.9737,\quad |V_{cb}|=0.041,\quad \tau^{}_{B_c}=0.510\ {\rm ps},\nonumber \\
&f_\pi=130.2\ {\rm MeV},\quad m_W=80.379\ {\rm GeV},\quad \bar{m}_c=1.27^{+0.02}_{-0.02}\ {\rm GeV},\nonumber\\
&\bar{m}_b=4.18^{+0.03}_{-0.02}\ {\rm GeV},\quad|R_{bc}(0)|^2=1.642\ {\rm GeV}^3,\quad |R_{cc}^\prime(0)|^2=0.075\ {\rm GeV}^5.
\end{align}
Here, $\bar{m}_Q$ denotes the $\overline{\rm MS}$ mass of heavy quark $Q$, $|R_{bc}(0)|$ and $|R_{cc}^\prime(0)|$ are evaluated in the QCD-motivated Buchm\"{o}ller-Tye potential \cite{Eichten:1995ch}.
At NLO, the pole mass $m_Q$ can be obtained through
\begin{equation}
m_Q=\bar{m}_Q\Big(1+C_F\frac{\alpha_s(\bar{m}_Q)}{\pi}\Big),
\end{equation}
where $C_F=4/3$ is the color factor. Then we have $m_b=4.60^{+0.03}_{-0.02}\ {\rm GeV}$ and $m_c=1.49^{+0.02}_{-0.02}\ {\rm GeV}$.

The two-loop formula \cite{Deur:2016tte}
\begin{equation}
\frac{4\pi}{\alpha_s(\mu^2)}-\frac{\beta_1}{\beta_0}\ln\left(\frac{4\pi}{\alpha_s(\mu^2)}+\frac{\beta_1}{\beta_0}\right) =\frac{4\pi}{\alpha_s(\mu_0^2)}-\frac{\beta_1}{\beta_0}\ln\left(\frac{4\pi}{\alpha_s(\mu_0^2)} +\frac{\beta_1}{\beta_0}\right)+\beta_0\ln\frac{\mu^2}{\mu_0^2}
\end{equation}
of the running coupling constant is employed in our calculation,
in which $\beta_1=(34/3)C_A^2-4C_FT_Fn_f-(20/3)C_AT_Fn_f$.
Here we adopt $n_f=4$, $\alpha_s(\mu_0^2)=0.184$ with $\mu_0=9.46$ GeV \cite{Brambilla:2007cz} as the initial scale.

For the parametrization of LCDA, it is convenient and conventional to expand it as:
\begin{equation}
\phi_{\pi}(x,\mu)=6x(1-x)\bigg(1+\sum_{n=1}^\infty a_{n}(\mu)C_{n}^{3/2}(1-2x)\bigg),
\label{eq_LCDA}
\end{equation}
where $C_n^{3/2}$ are the Gegenbauer polynomials
\begin{align}
&C_0^{3/2}(x)=1,\quad C_1^{3/2}(x)=3x,\quad C_2^{3/2}(x)=\frac{3}{2}(5x^2-1),\nonumber \\
&C_3^{3/2}(x)=\frac{5}{2}(7x^2-3),\quad C_4^{3/2}(x)=\frac{15}{8}(21x^4-14x^2+1),\cdots
\end{align}
Here we ignore the evolutionary effects, and take \cite{Braun:1989iv,Ball:1998je,Lu:2002ny}
\begin{equation}
a_2=0.44,\quad a_4=0.25,\quad {\rm others}=0.
\label{eq_apara}
\end{equation}
For convenience, we also introduce the moments
\begin{align}
&M_i^{(j)}(n)=\int_0^1 6x(1-x)C_n^{3/2}(1-2x)T_i^{(j)}(x),\\
&\tilde{M}_i^{(j)}(n)=\int_0^1 6x(1-x)C_n^{3/2}(1-2x)\tilde{T}_i^{(j)}(x).
\end{align}

\begin{figure}
\includegraphics[width=0.5\textwidth]{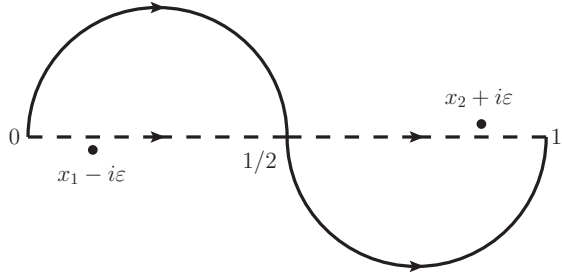}
\caption{The integral path used in numerical calculation.}
\label{fig_path}
\end{figure}

As has been discussed in Sec. \ref{sec_ana}, the poles $x_1$ and $x_2$ will not spoil the integral of Eq. \eqref{eq_factorization}.
However, in numerical calculations, it is inefficient to integrate along the real axis, as one has to approach the poles at a distance of order $|i\varepsilon|$.
Hence, in our calculations, we deform the integration path from real axis to two conjoined semicircles, as shown in Fig. \ref{fig_path}.
The corresponding analytic continuation of $T_i^{(1)}(x)$ is also performed, which is a bit of a tedious task.

\begin{table}
\caption{Numerical results for $M_i^{(j)}$ and $\tilde{M}_i^{(j)}$. Here $m_b=4.60$ GeV, $m_c=1.49$ GeV.}
\centering
{\scriptsize
  \begin{tabular}{p{1cm}<{\raggedright} p{2.3cm}<{\centering} p{2.3cm}<{\centering}p{2.3cm}<{\centering}p{2.3cm}<{\centering}p{2.3cm}<{\centering}p{2.3cm}<{\centering}}
 \toprule[2pt]
 \multicolumn{7}{c}{$M_i^{(j)}(n)$ of $B_c^+\to \chi_{c0}\pi^+$, in unit of $10^4$} \\
 \hline
       & $M_0^{(0)}(n)$ & $M_0^{(1)}(n)$ & $\tilde{M}_0^{(1)}(n)$ & $M_8^{(0)}(n)$ & $M_8^{(1)}(n)$ & $\tilde{M}_8^{(1)}(n)$ \\
 $n=0$ & $-51.1$ & $-607+3.15i$  & $-105-1.07i$   & $-1.03+0.716i$  & $51.9+29.6i$  & $14.4 + 1.85 i$ \\
 $n=1$ & $0$     & $14.4-1.44i$  & $0.759-0.799i$ & $-0.506+0.532i$ & $4.00+56.6i$  & $-1.76 + 1.85 i$ \\
 $n=2$ & $0$     & $11.3-1.01i$  & $-1.36-2.38i$  & $0.908+1.58i$   & $16.6+15.5i$  & $3.61 + 6.29 i$ \\
 $n=3$ & $0$     & $1.98-28.9i$  & $-1.00-1.72i$  & $0.670+1.14i$   & $31.9 +7.20i$ & $2.90 + 4.95 i$ \\
 $n=4$ & $0$     & $0.723-18.8i$ & $-3.02+1.67i$  & $2.01-1.11i$    & $20.9 -16.8i$ & $9.27 - 5.12 i$ \\
 \bottomrule[2pt]
  \multicolumn{7}{c}{$M_i^{(j)}(n)$ of $B_c^+\to \chi_{c1}\pi^+$, in unit of $10^4$} \\
 \hline
       & $M_0^{(0)}(n)$ & $M_0^{(1)}(n)$ & $\tilde{M}_0^{(1)}(n)$ & $M_8^{(0)}(n)$ & $M_8^{(1)}(n)$ & $\tilde{M}_8^{(1)}(n)$ \\
 $n=0$ & $-4.82$&$-54.2 + 3.01 i$&$-10.9 + 0.708 i$&$0.612 - 0.472 i$&$12.1 - 6.23 i$&$3.19 - 1.22 i$ \\
 $n=1$ & $0$&$-5.22 - 9.86 i$&$0.233 - 1.60 i$&$-0.155 + 1.07 i$&$-0.518 + 18.6 i$&$-0.538 + 3.71 i$ \\
 $n=2$ & $0$&$10.2 + 0.556 i$&$1.03 + 1.33 i$&$-0.686 - 0.888 i$&$-6.57 - 18.6 i$&$-2.73 - 3.53 i$ \\
 $n=3$ & $0$&$-15.0 + 13.3 i$&$-2.12 - 0.208 i$&$1.41 + 0.139 i$&$23.2 + 4.98 i$&$6.12 + 0.600 i$ \\
 $n=4$ & $0$&$-2.72 - 16.0 i$&$1.57 - 1.39 i$&$-1.04 + 0.923 i$&$-24.9 + 3.93 i$&$-4.81 + 4.25 i$ \\
 \bottomrule[2pt]
   \multicolumn{7}{c}{$M_i^{(j)}(n)$ of $B_c^+\to \chi_{c2}\pi^+$, in unit of $10^4$} \\
 \hline
       & $M_0^{(0)}(n)$ & $M_0^{(1)}(n)$ & $\tilde{M}_0^{(1)}(n)$ & $M_8^{(0)}(n)$ & $M_8^{(1)}(n)$ & $\tilde{M}_8^{(1)}(n)$ \\
$n=0$&$47.6$&$554 - 3.59 i$&$99.4 - 0.253 i$&$-0.0790 + 0.169 i$&$-54.3 - 16.7 i$&$-16.1 + 0.436 i$\\
$n=1$&$0$&$-4.06 + 5.58 i$&$-1.09 + 1.28 i$&$0.730 - 0.853 i$&$-5.03 - 40.8 i$&$2.53 - 2.96 i$\\
$n=2$&$0$&$-12.6 - 10.4 i$&$-0.649 + 0.0235 i$&$0.433 - 0.0157 i$&$-1.64 + 0.456 i$&$1.72 - 0.0622 i$\\
$n=3$&$0$&$4.30 + 7.99 i$&$1.62 + 2.45 i$&$-1.08 - 1.63 i$&$-13.3 - 5.67 i$&$-4.67 - 7.06 i$\\
$n=4$&$0$&$-16.6 + 22.6 i$&$0.349 + 1.11 i$&$-0.232 - 0.741 i$&$-2.51 + 2.35 i$&$-1.07 - 3.41 i$\\
 \bottomrule[2pt]
    \multicolumn{7}{c}{$M_i^{(j)}(n)$ of $B_c^+\to h_{c}\pi^+$, in unit of $10^4$} \\
 \hline
       & $M_0^{(0)}(n)$ & $M_0^{(1)}(n)$ & $\tilde{M}_0^{(1)}(n)$ & $M_8^{(0)}(n)$ & $M_8^{(1)}(n)$ & $\tilde{M}_8^{(1)}(n)$ \\
$n=0$&$66.7$&$795 - 5.37 i$&$138 + 0.484 i$&$0.916 - 0.322 i$&$-78.3 - 40.7 i$&$-19.9 - 0.833 i$\\
$n=1$&$0$&$-12.2 + 1.28 i$&$-0.529 + 1.02 i$&$0.353 - 0.677 i$&$-12.2 - 68.3 i$&$1.23 - 2.35 i$\\
$n=2$&$0$&$-12.7 - 0.513 i$&$-0.304 + 2.69 i$&$0.202 - 1.80 i$&$-8.48 - 5.48 i$&$0.804 - 7.13 i$\\
$n=3$&$0$&$-9.27 + 20.7 i$&$1.31 + 1.09 i$&$-0.876 - 0.727 i$&$-37.0 - 2.78 i$&$-3.79 - 3.15 i$\\
$n=4$&$0$&$2.04 + 21.1 i$&$4.31 + 1.25 i$&$-2.88 - 0.836 i$&$-7.85 - 0.762 i$&$-13.2 - 3.85 i$\\
 \bottomrule[2pt]
  \end{tabular}
}
\label{tab_Mr}
\end{table}

The numerical results for $M_i^{(j)}$ and $\tilde{M}_i^{(j)}$ are shown in Table \ref{tab_Mr}.
It can be seen that the values of $M_0^{(0)}(0)$, $M_0^{(1)}(0)$, and $\tilde{M}_0^{(1)}(0)$ are more significant than others.
These significant moments corresponding to the first term of Eq. \eqref{eq_LCDA}, which depicts the asymptotic behavior of LCDA: $\phi_\pi(x,\mu\to \infty)=6x(1-x)$.
Hence the uncertainty, caused either by ignoring the evolutionary effect of $a_n$ or by taking another parametrization scheme, is unimportant to the final decay width.

\subsection{Branching ratios}
In the calculation of branching ratios, one meets a considerable freedom in the choice of renormalization scale $\mu$,
because the meaning of $\mu$ is not quite evident.
By convention, there are three reasonable schemes:
\begin{enumerate}
\item Set $\mu$ to be some typical virtuality of internal particles. For example, the virtuality of gluons of Fig. \ref{fig_Feytree}(a)$\sim$(b) lead to $\mu= 1.1$ GeV; the largest virtuality of internal particles of Fig. \ref{fig_Feytree}(a)$\sim$(b) leads to  $\mu= 4.0$ GeV.
\item Relate $\mu$ to kinematic variables of final state particles. For example, the three-momentum of final state charmonium or pion in the $B_c^+$ rest frame lead to $\mu=|\vec{k}_2|= 2.3$ GeV.
\item Determine $\mu$ through scale-setting procedure, like the Brodsky-Lepage-Mackenzie (BLM) method \cite{Brodsky:1982gc,Brodsky:2011ta}, where the quark vacuum-polarization corrections are re-summed into the running coupling.
In our case, the BLM scale can be acquired by setting $\mu$ such it that kills the $n_f$ terms.
We obtain $\mu_{\rm BLM}=0.38$ GeV for $\chi_{c0}$ channel, $\mu_{\rm BLM}=0.47$ GeV for $\chi_{c1}$ channel, $\mu_{\rm BLM}=0.34$ GeV for $\chi_{c2}$ and $h_c$ channels.
\end{enumerate}
Since the BLM scales here are close to the perturbative QCD scale $\Lambda_{\rm QCD}$, we will not use them in following evaluation.
Instead, we will vary $\mu$ in a range that covers most reasonable choices of scale.

\begin{figure}
\centering
\subfigure{
\includegraphics[width=0.9\textwidth]{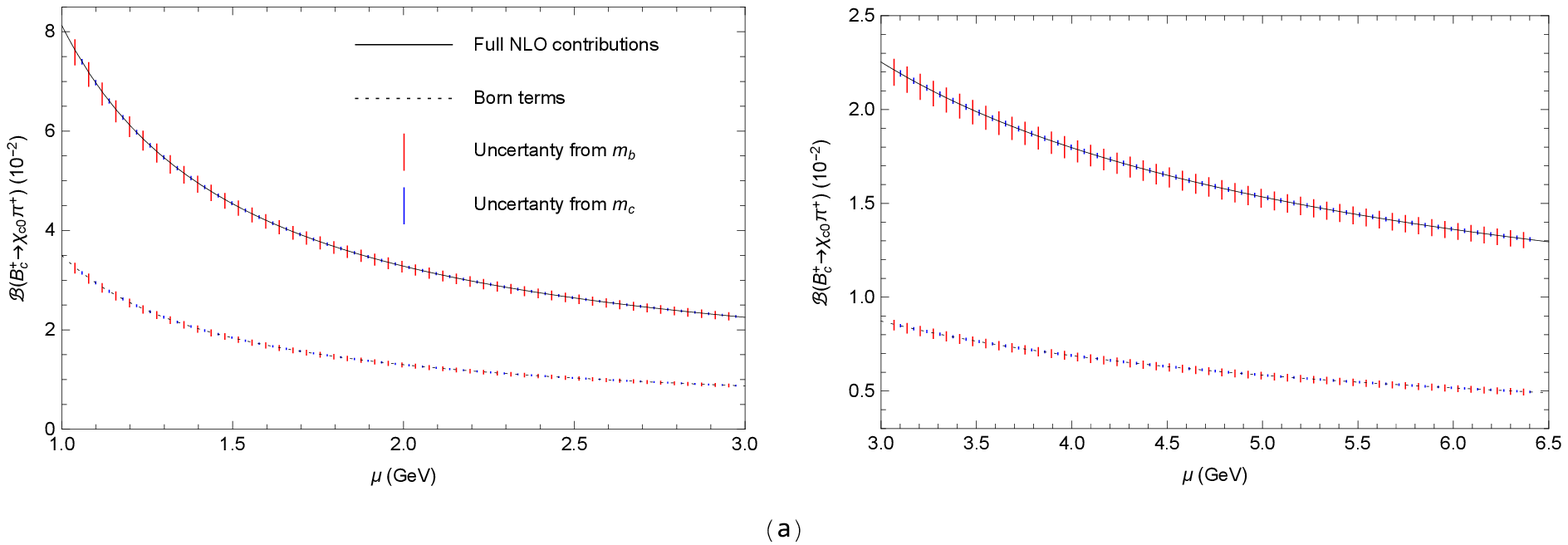}}
\subfigure{
\includegraphics[width=0.9\textwidth]{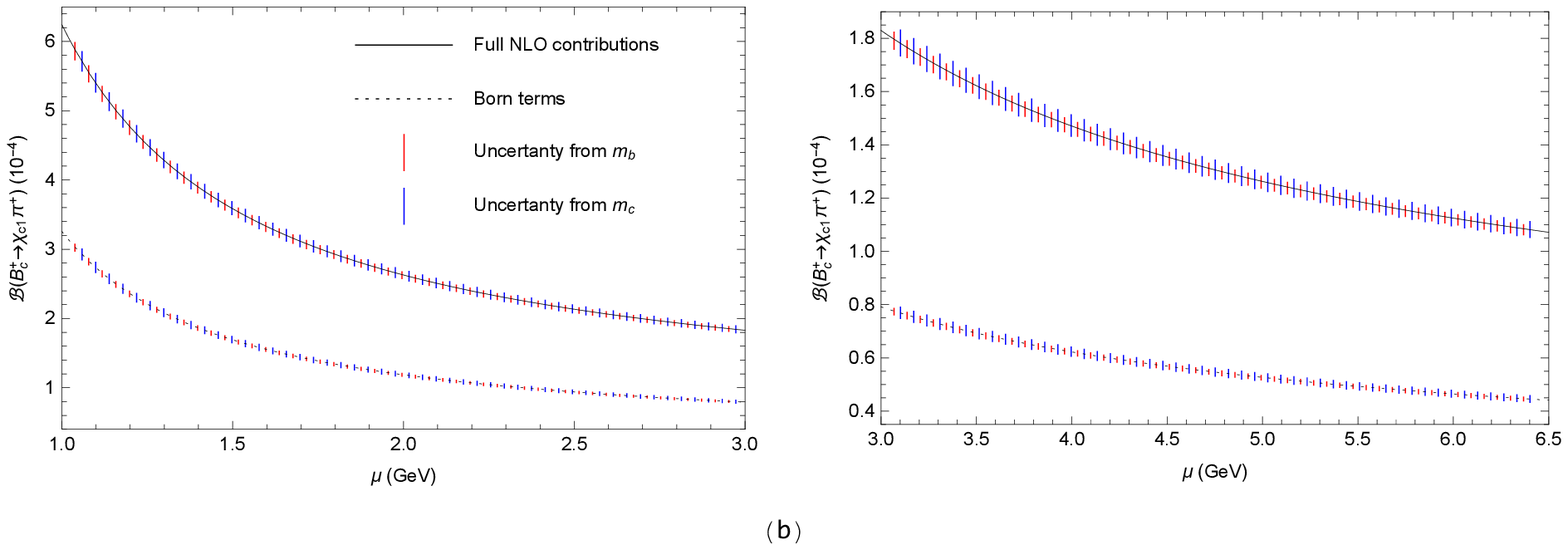}}
\subfigure{
\includegraphics[width=0.9\textwidth]{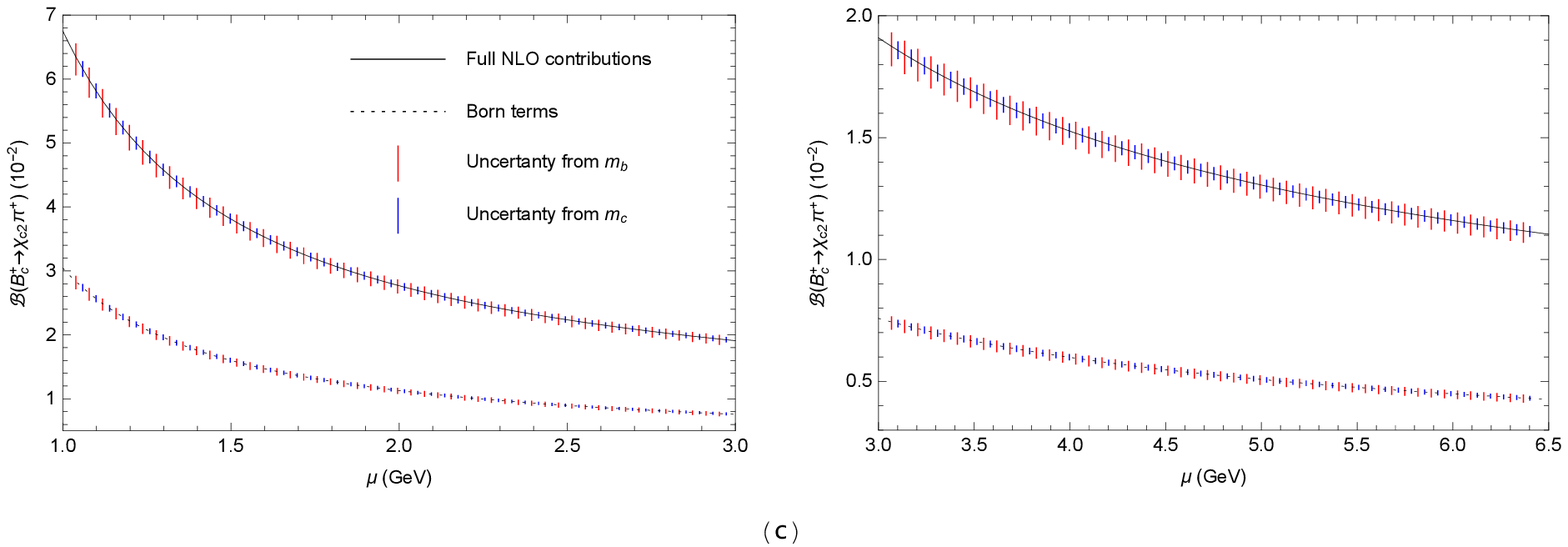}}
\subfigure{
\includegraphics[width=0.9\textwidth]{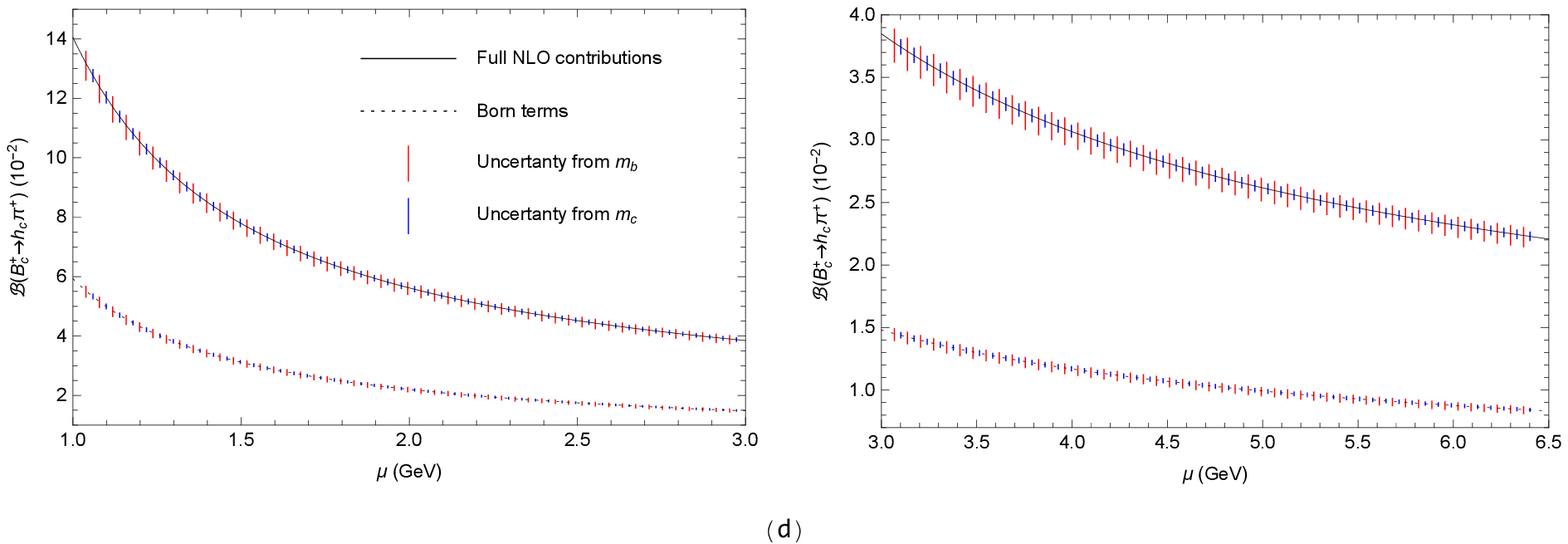}}
\caption{The LO and NLO branching ratio versus running renormalization scale $\mu$.
(a) for $B_c^+\to \chi_{c0}\pi^+$, (b) for $B_c^+\to \chi_{c1}\pi^+$, (c) for $B_c^+\to \chi_{c2}\pi^+$, (d) for $B_c^+\to h_c\pi^+$.
The plots in the left column corresponding to $1.0<\mu<3.0$, and in the right column corresponding to $3.0<\mu<6.5$.}
\label{fig_rchic0}
\end{figure}

The branching ratios can be obtained by employing Eq. \eqref{eq_Mtobr}.
The NLO branching ratios versus the running renormalization scale $\mu$ are exhibited in Fig. \ref{fig_rchic0},
where the contributions from the Born terms and the uncertainties induced by heavy quark masses are also shown.
It can be seen that, after including the NLO corrections, the branching ratios are enhanced by a factor (defined as the $K$ factor) of about $2.5$, and the theoretical uncertainties induced by the renormalization scale remain large, which may indicate that the contributions from missing higher orders (beyond NLO) are significant.

In Table \ref{tab_br}, we compare our branching ratios with those from other literatures.
Among them, the predictions of Ref. \cite{Kiselev:2001zb} are based on LO NRQCD calculation, with some technique details different from ours;
the predictions of Ref. \cite{Zhu:2017lwi} are also based on LO NRQCD, but with relativistic corrections.
It can be seen that our LO predictions are compatible with those of Refs. \cite{Kiselev:2001zb,Zhu:2017lwi}, but generally larger than those of other approaches.
Our NLO branching ratios, however, are lager than almost all other predictions.

\begin{table}
\caption{Comparison of the branching ratios of $B_c^+\to \chi_{cJ}(h_c)\pi^+$. Here we vary the renormalization scale $\mu$ from $|\vec{k}_2|/2$ to $2|\vec{k}_2|$.
The central values are obtained at $\mu=|\vec{k}_2|$.
The uncertainties are induced by $\mu$ and heavy quark masses.}
\centering
{\small
  \begin{tabular}{p{4.3cm}<{\raggedright} p{1.5cm}<{\centering} p{1.5cm}<{\centering}p{1cm}<{\centering}p{1cm}<{\centering}p{1cm}<{\centering}p{1cm}<{\centering}p{1cm}<{\centering}p{1cm}<{\centering}p{1cm}<{\centering}}
 \toprule[2pt]
  Branching ratios ($10^{-3}$)  & NLO & LO & \cite{Kiselev:2001zb} & \cite{Zhu:2017lwi} & \cite{Rui:2017pre} & \cite{Wang:2011jt} & \cite{Hernandez:2006gt} & \cite{Ivanov:2006ni} & \cite{Ebert:2010zu}     \\
    \hline
 $\mathcal{B}(B_c^+\to \chi_{c0}\pi^+)$            & $28^{+38}_{-13}$ & $11^{+16}_{-5.0}$  & $9.7$  & $6.5$ & $1.6$  & 0.3  & 0.26 & $0.55$ & 0.21 \\
 $10^2\times \mathcal{B}(B_c^+\to \chi_{c1}\pi^+)$ & $23^{+28}_{-10}$ & $10^{+16}_{-4.7}$  & $9.7$  & $6.4$ & $51$   & 2.2  & 0.14 & $6.8$  & 20 \\
 $\mathcal{B}(B_c^+\to \chi_{c2}\pi^+)$            & $24^{+31}_{-11}$ & $9.7^{+14}_{-4.5}$ & $8.5$  & $4.4$ & $4.0$  & 0.23 & 0.22 & $0.46$ & 0.38  \\
 $\mathcal{B}(B_c^+\to h_c\pi^+)      $            & $48^{+66}_{-22}$ & $19^{+28}_{-8.8}$  & $16$   & $9.7$ & $0.54$ & 1.0  & 0.53 & $1.1$  & 0.46  \\
 \bottomrule[2pt]
  \end{tabular}
}
\label{tab_br}
\end{table}

In Ref. \cite{Aaij:2016xas}, the LHCb Collaboration reported evidence for the decay $B_c^+\to \chi_{c0}\pi^+$ with a significance of 4.0 standard deviations.
The measured product of the ratio of cross sections and branching ratio is $\frac{\sigma(B_c^+)}{\sigma(B^+)}\times \mathcal{B}(B_c^+\to \chi_{c0}\pi^+)=(9.8^{+3.4}_{-3.0}({\rm stat})\pm0.8({\rm syst}))\times 10^{-6}$.
Combining this result with the measurement on $\frac{\sigma(B_c^+)}{\sigma(B^+)}\times \frac{\mathcal{B}(B_c^+\to J/\psi\pi^+)}{\mathcal{B}(B^+\to J/\psi K^+)}$ \cite{Aaij:2014ija} and the data for $\mathcal{B}(B^+\to J/\psi K^+)$ \cite{Zyla:2020zbs}, we obtain
\begin{equation}
\frac{\mathcal{B}(B_c^+\to \chi_{c0}\pi^+)}{\mathcal{B}(B_c^+\to J/\psi\pi^+)}=1.4_{-0.6}^{+0.8}.
\label{eq_meaR}
\end{equation}
In Ref. \cite{Qiao:2012hp}, the exclusive processes $B_c^+\to J/\psi\pi^+$ is investigated at the NLO QCD accuracy within the framework of the NRQCD factorization formalism.
By taking the same parameters as Ref. \cite{Qiao:2012hp}, we obtain
\begin{equation}
\frac{\mathcal{B}(B_c^+\to \chi_{c0}\pi^+)}{\mathcal{B}(B_c^+\to J/\psi\pi^+)}\bigg|_{\rm NLO}^{\rm NRQCD}=2.7_{-1.0}^{+1.7},
\end{equation}
which is compatible with Eq. \eqref{eq_meaR} with the uncertainties.
It should be remarked that the NRQCD predictions on both $\mathcal{B}(B_c^+\to \chi_{c0}\pi^+)$ and $\mathcal{B}(B_c^+\to J/\psi\pi^+)$ are larger than those of other approaches.
Further experimental studies, especially the measurement on the exact values of branching ratios, are needed to clarify the situation.

In the ``naive factorization'' approach, the exclusive two-body decay amplitude reduces to the product of a form factor and a decay constant.
The contributions of ``naive factorization'' come from the so-called ``factorizable'' diagrams, like Fig. \ref{fig_Feytree}(a)$\sim$(b) and Fig. \ref{fig_Feynlo}(a)$\sim$(b).
Our numerical estimation shows that the ``factorizable'' diagrams can mimic over $93\%$ contribution of the full calculation, which is sufficient in phenomenological study.
On the other hand, since the ``factorizable'' hard kernels are independent of $x$, the convolution in Eq. \eqref{eq_factorization} reduce to the integration of $\phi_\pi(x,\mu)$, which can be easily obtained.
The ``factorizable'' hard kernels are presented in the Appendix.

With the $M_i^{(j)}$ and $\tilde{M}_i^{(j)}$ listed in Table \ref{tab_Mr}, the branching ratios for $B_c^+\to \chi_{cJ}(h_c)K^+$ are also at hand.
Considering the fact that the contributions from ``factorizable'' diagrams are dominant, we have $\frac{\mathcal{B}(B_c^+\to \chi_{cJ}(h_c)K^+)}{\mathcal{B}(B_c^+\to \chi_{cJ}(h_c)\pi^+)}\approx \big(\frac{|V_{us}|f_K}{|V_{ud}|f_\pi}\big)^2\approx 0.0762$.

\section{Summary and conclusions}
In this work, we investigate the $B_c$ meson exclusive decays to a $P$-wave charmonium and a pion at the NLO QCD accuracy in the framework of the NRQCD factorization formalism.
Numerical results show that after including the NLO corrections, the branching ratios are enhanced by a factor of about $2.5$.
The renormalization scale dependence, however, is still large at NLO, which may indicate significant contributions from higher-order terms beyond NLO.

Although the exact values of $\mathcal{B}(B_c^+\to \chi_{cJ}(h_c)\pi^+)$ have not been measured yet,
 the quantity $\frac{\mathcal{B}(B_c^+\to \chi_{c0}\pi^+)}{\mathcal{B}(B_c^+\to J/\psi\pi^+)}$ can be extracted from relevant measurements \cite{Zyla:2020zbs,Aaij:2016xas,Aaij:2014ija}.
With the prediction of $\mathcal{B}(B_c^+\to J/\psi\pi^+)$ in Ref. \cite{Qiao:2012hp} and our prediction of $\mathcal{B}(B_c^+\to \chi_{c0}\pi^+)$, we find the predicted $\frac{\mathcal{B}(B_c^+\to \chi_{c0}\pi^+)}{\mathcal{B}(B_c^+\to J/\psi\pi^+)}$ is compatible with experimental data.
However, since the NRQCD predictions on both $\mathcal{B}(B_c^+\to \chi_{c0}\pi^+)$ and $\mathcal{B}(B_c^+\to J/\psi\pi^+)$ are generally larger than those of other approaches,
further experimental study is still needed to clarify this issue.

Last, we notice that the contributions from ``factorizable'' diagrams are dominant, hence the ``naive factorization'' approach works well in at least our phenomenological study.
Therefore we have $\frac{\mathcal{B}(B_c^+\to \chi_{cJ}(h_c)K^+)}{\mathcal{B}(B_c^+\to \chi_{cJ}(h_c)\pi^+)}\approx \big(\frac{|V_{us}|f_K}{|V_{ud}|f_\pi}\big)^2\approx 0.0762$.
\newpage
\vspace{0.0cm} {\bf Acknowledgments}

This work was supported in part by the National Key Research and Development Program of China under Contracts No. 2020YFA0406400,
and the National Natural Science Foundation of China (NSFC) under the Grants No. 11975236, No. 11635009, and  No. 12047553.

\section*{Appendix: ``factorizable'' hard kernels}
In this appendix, we present the ``factorizable'' hard kernels $T_{0,{\rm f}}^{(0)}$, $\tilde{T}_{0,{\rm f}}^{(1)}$ and $T_{0,{\rm f}}^{(1)}$.
The LO ``factorizable'' hard kernels $T_{0,{\rm f}}^{(0)}=T_{0}^{(0)}$, which have been presented in Eqs. \eqref{eq_Ttree1}\eqref{eq_Ttree3}\eqref{eq_Ttree5}\eqref{eq_Ttree7}.
The NLO ``factorizable'' hard kernels $\tilde{T}_{0,{\rm f}}^{(1)}=\frac{\beta_0}{4}T_{0}^{(0)}$, and $T^{(1)}_{0,{\rm f}}$ are as follows:
\begin{align}
T&_{0,{\rm f}}^{(1)}(\chi_{c0})=\Big\{\tfrac{4 (11 z-24)}{z^2-2 z-1}+(-\tfrac{4 (74 z-219)}{z^2-2 z-1}+\tfrac{839 z^2-4048 z+5525}{32 (z^3-4 z^2+3 z+4)}-\tfrac{4}{9 (z-3)}+\tfrac{458}{z-1}+\tfrac{8 (z-6)}{(z^2-2 z-1)^2}\nonumber \\
&-\tfrac{64 z}{(z^2-2 z-1)^3}-\tfrac{14}{z+1}-\tfrac{3 (7 z-17)}{2 (z^2-3 z+4)}-\tfrac{47023}{288 z}-\tfrac{20497}{24 z^2}-\tfrac{1019}{2 z^3}-\tfrac{762}{z^4}-\tfrac{224}{z^5}+\tfrac{480}{z^6}-\tfrac{384}{z^7}) f_2\nonumber \\
&+(\tfrac{3 (z-7)}{2 (z^2-3 z+4)}+\tfrac{839 z^2-4048 z+5525}{16 (z^3-4 z^2+3 z+4)}-\tfrac{8}{9 (z-3)}-\tfrac{106}{9 (z-2)}+\tfrac{70}{3 (z-2)^2}-\tfrac{870}{z-1}-\tfrac{976}{3 (2 z-1)}+\tfrac{224}{3 (2 z-1)^2}\nonumber \\
&+\tfrac{310}{9 (z+1)}-\tfrac{56}{3 (z+1)^2}+\tfrac{145865}{144 z}+\tfrac{12971}{12 z^2}+\tfrac{1865}{3 z^3}+\tfrac{512}{z^4}+\tfrac{664}{3 z^5}+\tfrac{480}{z^6}-\tfrac{384}{z^7}) f_3\nonumber \\
&+(\tfrac{-839 z^2+4048 z-5525}{32 (z^3-4 z^2+3 z+4)}+\tfrac{4}{9 (z-3)}-\tfrac{22}{z-2}+\tfrac{14}{(z-2)^2}+\tfrac{128}{z-1}+\tfrac{320}{3 (z-1)^2}-\tfrac{8 (74 z-219)}{z^2-2 z-1}+\tfrac{16 (z-6)}{(z^2-2 z-1)^2}\nonumber \\
&-\tfrac{128 z}{(z^2-2 z-1)^3}-\tfrac{28}{z+1}-\tfrac{12 (2 z-3)}{2 z^2-5 z+1}+\tfrac{48}{z^2-3 z+4}+\tfrac{142783}{288 z}-\tfrac{5197}{8 z^2}-\tfrac{3763}{6 z^3}-\tfrac{586}{z^4}-\tfrac{1336}{3 z^5}) f_4\nonumber \\
&+(\tfrac{8 (74 z-219)}{z^2-2 z-1}+\tfrac{128 z}{(z^2-2 z-1)^3}-\tfrac{734}{z-1}-\tfrac{320}{3 (z-1)^2}+\tfrac{1952}{3 (2 z-1)}-\tfrac{448}{3 (2 z-1)^2}-\tfrac{16 (z-6)}{(z^2-2 z-1)^2}+\tfrac{320}{3 (z+1)}\nonumber \\
&+\tfrac{112}{3 (z+1)^2}+\tfrac{6 (2 z-3)}{2 z^2-5 z+1}+\tfrac{12 (z-3)}{z^2-3 z+4}-\tfrac{308}{z}+\tfrac{4328}{3 z^2}+\tfrac{888}{z^3}+\tfrac{2000}{z^4}) f_5+(\tfrac{12 (z-3)}{z^2-3 z+4}-\tfrac{6}{z-1}\nonumber \\
&+\tfrac{6 (2 z-3)}{2 z^2-5 z+1}-\tfrac{12}{z}) f_6+(\tfrac{2 (29 z^2-143 z-210)}{11 (z^3-5 z^2-8 z-4)}+\tfrac{3967 z^2-18000 z+23101}{176 (z^3-4 z^2+3 z+4)}+\tfrac{14}{z+1}-\tfrac{669}{16 z}+\tfrac{83}{4 z^2}\nonumber \\
&-\tfrac{13}{z^3}-\tfrac{16}{z^4}) f_7+(\tfrac{6 (z-7)}{z^2-3 z+4}+\tfrac{1052}{9 (z-2)}-\tfrac{140}{3 (z-2)^2}-\tfrac{160}{z-1}+\tfrac{352}{9 (z+1)}-\tfrac{2}{z}+\tfrac{36}{z^2}-\tfrac{64}{z^3}+\tfrac{16}{z^4}\nonumber \\
&-\tfrac{64}{z^5}) f_8+(-\tfrac{28}{9 (z-2)^2}-\tfrac{12122}{27 (z+1)}-\tfrac{880}{9 (z+1)^2}-\tfrac{6}{z^2-3 z+4}+\tfrac{442}{z}-\tfrac{356}{z^2}+\tfrac{256}{z^3}-\tfrac{144}{z^4}+\tfrac{64}{z^5}\nonumber \\
&+\tfrac{188}{27 (z-2)}) f_9+(\tfrac{3 (5 z-3)}{2 (z^2-3 z+4)}-\tfrac{80}{9 (z-2)}+\tfrac{14}{3 (z-2)^2}-\tfrac{352}{9 (z+1)}+\tfrac{81}{2 z}-\tfrac{44}{z^2}+\tfrac{32}{z^3}-\tfrac{24}{z^4}) f_{10}\nonumber \\
&+(\tfrac{3 (7 z-17)}{2 (z^2-3 z+4)}-\tfrac{8}{z-2}+\tfrac{14}{(z-2)^2}-\tfrac{5}{2 z}-\tfrac{4}{z^2}-\tfrac{8}{z^4}) f_{11}+(-3 z+\tfrac{2}{z-1}-\tfrac{14}{z+1}-\tfrac{8}{z^2}-\tfrac{12}{z^3}+10\nonumber \\
&-\tfrac{2}{z}) f_{12}+(\tfrac{14}{z+1}-\tfrac{57}{z}-\tfrac{63}{z^2}-\tfrac{52}{z^3}-\tfrac{76}{z^4}-37+\tfrac{18}{z-1}) f_{13}+(-\tfrac{3 (5 z-3)}{z^2-3 z+4}-\tfrac{72}{5 (z+1)}\nonumber \\
&+\tfrac{36 (2 z-1)}{5 (z^2-2 z+2)}+\tfrac{3}{z}+3) f_{14}+(25 z+\tfrac{112}{z-1}-\tfrac{36}{z^2-2 z+2}-\tfrac{93}{z^2}+\tfrac{24}{z^3}-\tfrac{36}{z^4}+103-\tfrac{6}{z}) f_{15}+(\tfrac{2}{z}\nonumber \\
&+\tfrac{10}{z^2}-\tfrac{28}{z^3}+\tfrac{16}{z^4}+\tfrac{6}{z+1}) f_{16}+6 f_{17}+(\tfrac{12 (z+1)}{z^2-3 z+4}-3 z+\tfrac{6}{z}-\tfrac{16}{z^2}+\tfrac{4}{z^3}+1-\tfrac{32}{z+1}) f_{18}\nonumber \\
&+(-2 z^2-\tfrac{6}{z+1}+\tfrac{30}{z^2}+\tfrac{38}{z^3}-\tfrac{24}{z^4}-11-\tfrac{17}{z}) f_{19}+(-2 z^2-\tfrac{5 z}{2}+\tfrac{27}{2 z^2}-\tfrac{10}{z^3}+\tfrac{8}{z^4}-31+\tfrac{2}{z}) f_{20}\nonumber \\
&+(-\tfrac{32}{z}+\tfrac{32}{z^2}-\tfrac{32}{z^3}+\tfrac{16}{z^4}+6+\tfrac{32}{z+1}) f_{21}+(-\tfrac{29 z}{2}+\tfrac{16}{z-1}+\tfrac{89}{2 z^2}+\tfrac{4}{z^3}+\tfrac{16}{z^4}-51+\tfrac{16}{z}) f_{22}\nonumber \\
&+(\tfrac{10}{z}+\tfrac{22}{z^2}+\tfrac{36}{z^3}+\tfrac{16}{z^4}-\tfrac{2}{z-1}) f_{23}+(4 z^2-20 z+\tfrac{40}{z^2}+68-\tfrac{40}{z}) f_{24}+(\tfrac{72}{5 (z+1)}\nonumber \\
&-\tfrac{36 (2 z-1)}{5 (z^2-2 z+2)}) f_{25}+(\tfrac{36}{z^2-2 z+2}-\tfrac{72}{z^2-3 z+4}) f_{26}+\tfrac{152}{3 (z-2)}-\tfrac{56}{(z-2)^2}+\tfrac{814}{3 (z-1)}+\tfrac{224}{2 z-1}\nonumber \\
&-\tfrac{32 (z-2)}{(z^2-2 z-1)^2}+\tfrac{436}{3 (z+1)}-\tfrac{12 (5 z-9)}{z^2-3 z+4}-\tfrac{1690}{3 z}-\tfrac{892}{3 z^2}-\tfrac{116}{z^3}-\tfrac{1240}{3 z^4}+\tfrac{288}{z^5}-\tfrac{384}{z^6}\Big\}\tfrac{64\pi^2}{9},\\
T&_{0,{\rm f}}^{(1)}(\chi_{c1})=\Big\{(-\tfrac{8 (2 z-5)}{z^2-2 z-1}+\tfrac{32 (2 z+1)}{(z^2-2 z-1)^2}+\tfrac{8}{z-3}+\tfrac{1370}{3 (z-2)}-\tfrac{164}{z-1}-\tfrac{8}{3 (z+1)}-\tfrac{118}{z}-\tfrac{360}{z^2}+\tfrac{8}{z^3}\nonumber \\
&+\tfrac{76}{z^4}-\tfrac{64}{z^5}) f_2+(\tfrac{341}{3 (z-2)}-\tfrac{172}{z-1}-\tfrac{304}{3 (2 z-1)}+\tfrac{32}{(2 z-1)^2}+\tfrac{88}{z+1}-\tfrac{16}{(z+1)^2}+\tfrac{121}{z}+\tfrac{506}{3 z^2}+\tfrac{508}{3 z^3}\nonumber \\
&+\tfrac{68}{z^4}-\tfrac{64}{z^5}+\tfrac{16}{z-3}) f_3+(-\tfrac{16 (2 z-5)}{z^2-2 z-1}+\tfrac{64 (2 z+1)}{(z^2-2 z-1)^2}-\tfrac{8}{z-3}+\tfrac{1345}{3 (z-2)}-\tfrac{32}{z-1}+\tfrac{64}{3 (z-1)^2}-\tfrac{16}{3 (z+1)}\nonumber \\
&+\tfrac{24 (2 z-1)}{2 z^2-5 z+1}-\tfrac{123}{z}-\tfrac{890}{3 z^2}-\tfrac{484}{3 z^3}+\tfrac{8}{z^4}) f_4+(\tfrac{16 (2 z-5)}{z^2-2 z-1}-\tfrac{460}{z-2}-\tfrac{244}{z-1}-\tfrac{64}{3 (z-1)^2}+\tfrac{608}{3 (2 z-1)}\nonumber \\
&-\tfrac{64}{(2 z-1)^2}-\tfrac{64 (2 z+1)}{(z^2-2 z-1)^2}-\tfrac{544}{3 (z+1)}+\tfrac{32}{(z+1)^2}-\tfrac{12 (2 z-1)}{2 z^2-5 z+1}+\tfrac{380}{z}+\tfrac{696}{z^2}) f_5+(-\tfrac{12 (2 z-1)}{2 z^2-5 z+1}+\tfrac{16}{z-2}\nonumber \\
&-\tfrac{12}{z-1}+\tfrac{8}{z+1}) f_6+(\tfrac{8}{3 (z+1)}-\tfrac{11}{2 z}-\tfrac{7}{z^2}+\tfrac{2}{z^3}+\tfrac{41}{6 (z-2)}) f_7+(-2 z-\tfrac{7}{3 (z-2)}+\tfrac{4}{z-1}+\tfrac{28}{3 (z+1)}\nonumber \\
&+2-\tfrac{11}{z}) f_{12}+(-8 z-\tfrac{211}{6 (z-2)}+\tfrac{36}{z-1}-\tfrac{28}{3 (z+1)}-\tfrac{15}{2 z}-\tfrac{41}{z^2}-18) f_{13}+(-6 z-\tfrac{12}{z-2}) f_{14}\nonumber \\
&+(-6 z+\tfrac{83}{2 (z-2)}-\tfrac{36}{z-1}-\tfrac{31}{2 z}-\tfrac{3}{z^2}+18) f_{15}+(\tfrac{28}{3 (z+1)}-\tfrac{13}{z}+\tfrac{6}{z^2}-\tfrac{13}{3 (z-2)}) f_{16}+(-\tfrac{16}{z+1}\nonumber \\
&-\tfrac{8}{z-2}) f_{17}+(12 z+\tfrac{97}{3 (z-2)}+\tfrac{32}{3 (z+1)}+2+\tfrac{1}{z}) f_{18}+(4 z^2-31 z-\tfrac{13}{6 (z-2)}-\tfrac{28}{3 (z+1)}+\tfrac{31}{2 z}\nonumber \\
&-\tfrac{9}{z^2}+38) f_{19}+(-4 z^2-18 z-\tfrac{461}{4 (z-2)}+\tfrac{1}{4 z}-\tfrac{3}{2 z^2}-57) f_{20}+(-\tfrac{32}{3 (z+1)}-\tfrac{6}{z}+\tfrac{4}{z^2}\nonumber \\
&-\tfrac{22}{3 (z-2)}) f_{21}+(-8 z^2-25 z-\tfrac{753}{4 (z-2)}+\tfrac{34}{z-1}+\tfrac{5}{4 z}+\tfrac{21}{2 z^2}-35) f_{22}+(-\tfrac{4}{z-1}+\tfrac{13}{z}+\tfrac{10}{z^2}\nonumber \\
&-\tfrac{17}{z-2}) f_{23}+(8 z^2+8 z+\tfrac{60}{z-2}+32+\tfrac{4}{z}) f_{24}-\tfrac{500}{3 (z-1)}+\tfrac{96}{2 z-1}+\tfrac{16}{z^2-2 z-1}+\tfrac{24}{z+1}+\tfrac{476}{3 z}\nonumber \\
&+\tfrac{568}{3 z^2}+\tfrac{32}{z^3}-\tfrac{64}{z^4}\Big\}\tfrac{32}{3} \sqrt{\tfrac{2}{3}} \pi ^2,\\
T&_{0,{\rm f}}^{(1)}(\chi_{c2})=\Big\{\tfrac{18 (3 z-5)}{z^2-3 z+4}+\tfrac{4 z}{z^2-2 z-1}+(-\tfrac{3 (65 z-231)}{32 (z^2-3 z+4)}-\tfrac{9}{2 (z-2)}-\tfrac{12}{z-1}+\tfrac{723}{32 z}+\tfrac{141}{8 z^2}-\tfrac{9}{2 z^3}\nonumber \\
&-\tfrac{18}{z^4}) f_1+(\tfrac{-839 z^2+4048 z-5525}{32 (z^3-4 z^2+3 z+4)}+\tfrac{8 (11 z+6)}{(z^2-2 z-1)^2}+\tfrac{64 z}{(z^2-2 z-1)^3}+\tfrac{4}{9 (z-3)}-\tfrac{1411}{2 (z-2)}+\tfrac{1017}{(z-2)^2}+\tfrac{490}{z-1}\nonumber \\
&+\tfrac{4 (80 z-243)}{z^2-2 z-1}+\tfrac{28}{z+1}+\tfrac{3 (37 z-163)}{16 (z^2-3 z+4)}-\tfrac{32735}{288 z}+\tfrac{14899}{24 z^2}-\tfrac{79}{2 z^3}+\tfrac{870}{z^4}+\tfrac{188}{z^5}-\tfrac{576}{z^6}+\tfrac{384}{z^7}) f_2\nonumber \\
&+(\tfrac{-839 z^2+4048 z-5525}{16 (z^3-4 z^2+3 z+4)}+\tfrac{8}{9 (z-3)}+\tfrac{3905}{18 (z-2)}-\tfrac{481}{3 (z-2)^2}+\tfrac{186}{z-1}-\tfrac{336}{2 z-1}+\tfrac{160}{3 (2 z-1)^2}-\tfrac{2884}{9 (z+1)}\nonumber \\
&-\tfrac{424}{3 (z+1)^2}-\tfrac{21 (17 z-23)}{16 (z^2-3 z+4)}+\tfrac{6313}{36 z}-\tfrac{1723}{6 z^2}-\tfrac{140}{3 z^3}-\tfrac{204}{z^4}-\tfrac{700}{3 z^5}-\tfrac{576}{z^6}+\tfrac{384}{z^7}) f_3\nonumber \\
&+(\tfrac{8 (80 z-243)}{z^2-2 z-1}+\tfrac{16 (11 z+6)}{(z^2-2 z-1)^2}+\tfrac{839 z^2-4048 z+5525}{32 (z^3-4 z^2+3 z+4)}+\tfrac{128 z}{(z^2-2 z-1)^3}-\tfrac{4}{9 (z-3)}+\tfrac{137}{z-2}+\tfrac{484}{(z-2)^2}\nonumber \\
&+\tfrac{424}{z-1}-\tfrac{128}{3 (z-1)^2}+\tfrac{48}{z+1}+\tfrac{12 (2 z-3)}{2 z^2-5 z+1}+\tfrac{3 (27 z-157)}{16 (z^2-3 z+4)}-\tfrac{376657}{288 z}+\tfrac{623}{8 z^2}-\tfrac{1163}{6 z^3}+\tfrac{350}{z^4}\nonumber \\
&+\tfrac{1264}{3 z^5}) f_4+(-\tfrac{8 (80 z-243)}{z^2-2 z-1}-\tfrac{1988}{3 (z-2)}-\tfrac{28}{(z-2)^2}+\tfrac{146}{z-1}+\tfrac{128}{3 (z-1)^2}+\tfrac{672}{2 z-1}-\tfrac{320}{3 (2 z-1)^2}\nonumber \\
&-\tfrac{16 (11 z+6)}{(z^2-2 z-1)^2}-\tfrac{128 z}{(z^2-2 z-1)^3}+\tfrac{860}{3 (z+1)}+\tfrac{272}{3 (z+1)^2}-\tfrac{6 (2 z-3)}{2 z^2-5 z+1}-\tfrac{12 (z-3)}{z^2-3 z+4}+\tfrac{552}{z}-\tfrac{620}{3 z^2}\nonumber \\
&+\tfrac{192}{z^3}-\tfrac{2000}{z^4}) f_5+(-\tfrac{12 (z-3)}{z^2-3 z+4}+\tfrac{22}{3 (z-2)}-\tfrac{14}{(z-2)^2}+\tfrac{6}{z-1}+\tfrac{248}{3 (z+1)}+\tfrac{64}{(z+1)^2}-\tfrac{6 (2 z-3)}{2 z^2-5 z+1}-\tfrac{78}{z}\nonumber \\
&+\tfrac{18}{z^2}) f_6+(\tfrac{-3967 z^2+18000 z-23101}{176 (z^3-4 z^2+3 z+4)}-\tfrac{2 (29 z^2-143 z-210)}{11 (z^3-5 z^2-8 z-4)}-\tfrac{7}{4 (z-2)}+\tfrac{6}{(z-2)^2}-\tfrac{4}{z+1}+\tfrac{537}{16 z}-\tfrac{65}{4 z^2}\nonumber \\
&+\tfrac{10}{z^3}+\tfrac{10}{z^4}) f_7+(-\tfrac{6 (z-7)}{z^2-3 z+4}-\tfrac{1052}{9 (z-2)}+\tfrac{140}{3 (z-2)^2}+\tfrac{160}{z-1}-\tfrac{352}{9 (z+1)}+\tfrac{2}{z}-\tfrac{36}{z^2}+\tfrac{64}{z^3}-\tfrac{16}{z^4}\nonumber \\
&+\tfrac{64}{z^5}) f_8+(\tfrac{28}{9 (z-2)^2}+\tfrac{12122}{27 (z+1)}+\tfrac{880}{9 (z+1)^2}+\tfrac{6}{z^2-3 z+4}-\tfrac{442}{z}+\tfrac{356}{z^2}-\tfrac{256}{z^3}+\tfrac{144}{z^4}-\tfrac{64}{z^5}\nonumber \\
&-\tfrac{188}{27 (z-2)}) f_9+(-\tfrac{3 (5 z-3)}{2 (z^2-3 z+4)}+\tfrac{80}{9 (z-2)}-\tfrac{14}{3 (z-2)^2}+\tfrac{352}{9 (z+1)}-\tfrac{81}{2 z}+\tfrac{44}{z^2}-\tfrac{32}{z^3}+\tfrac{24}{z^4}) f_{10}\nonumber \\
&+(-\tfrac{3 (7 z-17)}{2 (z^2-3 z+4)}+\tfrac{8}{z-2}-\tfrac{14}{(z-2)^2}+\tfrac{5}{2 z}+\tfrac{4}{z^2}+\tfrac{8}{z^4}) f_{11}+(-\tfrac{4}{(z-2)^2}-\tfrac{2}{z-1}+\tfrac{2}{3 (z+1)}+\tfrac{43}{2 z}-\tfrac{13}{z^2}\nonumber \\
&+\tfrac{12}{z^3}-4-\tfrac{19}{6 (z-2)}) f_{12}+(-\tfrac{2}{(z-2)^2}-\tfrac{18}{z-1}-\tfrac{2}{3 (z+1)}-\tfrac{189}{4 z}+\tfrac{153}{2 z^2}-\tfrac{11}{z^3}+\tfrac{76}{z^4}-14-\tfrac{457}{12 (z-2)}) f_{13}\nonumber \\
&+(\tfrac{3 (5 z-3)}{z^2-3 z+4}-\tfrac{9}{z-2}-\tfrac{9}{(z-2)^2}-\tfrac{9 (z-1)}{z^2-2 z+2}-\tfrac{3}{z}-3) f_{14}+(\tfrac{9 (z+1)}{z^2-2 z+2}-22 z+\tfrac{225}{4 (z-2)}+\tfrac{15}{(z-2)^2}\nonumber \\
&-\tfrac{16}{z-1}+\tfrac{183}{4 z}+\tfrac{21}{2 z^2}-\tfrac{33}{z^3}+\tfrac{36}{z^4}+29) f_{15}+(-\tfrac{3}{(z-2)^2}-\tfrac{38}{z+1}+\tfrac{37}{z}-\tfrac{43}{z^2}+\tfrac{40}{z^3}-\tfrac{16}{z^4}-\tfrac{1}{z-2}) f_{16}\nonumber \\
&+(-\tfrac{8}{(z-2)^2}+\tfrac{8}{3 (z+1)}-\tfrac{128}{(z+1)^2}+\tfrac{36}{z}-6-\tfrac{8}{3 (z-2)}) f_{17}+(-\tfrac{12 (z+1)}{z^2-3 z+4}+\tfrac{29}{6 (z-2)}+\tfrac{8}{(z-2)^2}\nonumber \\
&+\tfrac{104}{3 (z+1)}-\tfrac{33}{2 z}+\tfrac{19}{z^2}-\tfrac{4}{z^3}+5) f_{18}+(2 z^2+36 z+\tfrac{769}{4 (z-2)}+\tfrac{447}{4 (z-2)^2}+\tfrac{38}{z+1}+\tfrac{83}{4 z}+\tfrac{27}{4 z^2}-\tfrac{56}{z^3}\nonumber \\
&+\tfrac{24}{z^4}+17) f_{19}+(2 z^2+37 z-\tfrac{819}{4 (z-2)}-\tfrac{276}{(z-2)^2}+\tfrac{145}{4 z}-\tfrac{24}{z^2}+\tfrac{16}{z^3}-\tfrac{8}{z^4}-14) f_{20}\nonumber \\
&+(-\tfrac{44}{(z-2)^2}-\tfrac{104}{3 (z+1)}+\tfrac{41}{z}-\tfrac{50}{z^2}+\tfrac{44}{z^3}-\tfrac{16}{z^4}-6-\tfrac{127}{3 (z-2)}) f_{21}+(z-\tfrac{105}{z-2}-\tfrac{507}{4 (z-2)^2}-\tfrac{16}{z-1}\nonumber \\
&-\tfrac{5}{2 z}-\tfrac{187}{4 z^2}+\tfrac{8}{z^3}-\tfrac{16}{z^4}-39) f_{22}+(-\tfrac{12}{(z-2)^2}+\tfrac{2}{z-1}-\tfrac{10}{z}-\tfrac{16}{z^2}-\tfrac{24}{z^3}-\tfrac{16}{z^4}+\tfrac{6}{z-2}) f_{23}\nonumber \\
&+(-4 z^2-52 z-\tfrac{306}{z-2}-\tfrac{162}{(z-2)^2}-\tfrac{22}{z^2}-104+\tfrac{34}{z}) f_{24}+(\tfrac{9 (z-1)}{z^2-2 z+2}-\tfrac{9}{4 (z-2)}-\tfrac{9}{4 (z-2)^2}-\tfrac{27}{4 z}\nonumber \\
&-\tfrac{9}{4 z^2}+\tfrac{18}{z^3}-\tfrac{9}{z^4}) f_{25}+(-\tfrac{9 (z+1)}{z^2-2 z+2}+\tfrac{9 (93 z+85)}{64 (z^2-3 z+4)}-\tfrac{9}{z-2}-\tfrac{27}{4 (z-2)^2}+\tfrac{315}{64 z}+\tfrac{9}{16 z^2}-\tfrac{9}{4 z^3}) f_{26}\nonumber \\
&-\tfrac{152}{3 (z-2)}+\tfrac{56}{(z-2)^2}-\tfrac{550}{3 (z-1)}+\tfrac{160}{2 z-1}+\tfrac{32 (z-2)}{(z^2-2 z-1)^2}-\tfrac{292}{3 (z+1)}+\tfrac{580}{3 z}+\tfrac{202}{3 z^2}+\tfrac{160}{z^3}+\tfrac{1096}{3 z^4}\nonumber \\
&-\tfrac{384}{z^5}+\tfrac{384}{z^6}\Big\}\tfrac{64 \sqrt{2} \pi ^2}{9},\\
T&_{0,{\rm f}}^{(1)}(h_c)=\Big\{-\tfrac{4 (10 z-19)}{z^2-2 z-1}+(\tfrac{-683 z^2+3216 z-4289}{32 (z^3-4 z^2+3 z+4)}+\tfrac{40}{27 (z-3)}-\tfrac{794}{3 (z-2)}-\tfrac{272}{z-1}+\tfrac{4 (127 z-206)}{z^2-2 z-1}\nonumber \\
&-\tfrac{8 (12 z+7)}{(z^2-2 z-1)^2}-\tfrac{64 (2 z+1)}{(z^2-2 z-1)^3}+\tfrac{26}{3 (z+1)}-\tfrac{37271}{864 z}+\tfrac{52447}{72 z^2}+\tfrac{1117}{6 z^3}+\tfrac{634}{z^4}+\tfrac{296}{z^5}-\tfrac{544}{z^6}\nonumber \\
&+\tfrac{384}{z^7}) f_2+(\tfrac{-683 z^2+3216 z-4289}{16 (z^3-4 z^2+3 z+4)}+\tfrac{80}{27 (z-3)}-\tfrac{1301}{9 (z-2)}-\tfrac{10}{(z-2)^2}+\tfrac{418}{z-1}+\tfrac{448}{9 (2 z-1)}-\tfrac{16}{3 (2 z-1)^2}\nonumber \\
&+\tfrac{80}{3 (z+1)}+\tfrac{8}{3 (z+1)^2}-\tfrac{125399}{432 z}-\tfrac{15257}{36 z^2}-\tfrac{147}{z^3}-\tfrac{1496}{3 z^4}-\tfrac{688}{3 z^5}-\tfrac{544}{z^6}+\tfrac{384}{z^7}) f_3+(\tfrac{8 (127 z-206)}{z^2-2 z-1}\nonumber \\
&+\tfrac{683 z^2-3216 z+4289}{32 (z^3-4 z^2+3 z+4)}-\tfrac{40}{27 (z-3)}-\tfrac{1615}{3 (z-2)}-\tfrac{6}{(z-2)^2}+\tfrac{1052}{3 (z-1)}-\tfrac{32}{3 (z-1)^2}-\tfrac{16 (12 z+7)}{(z^2-2 z-1)^2}\nonumber \\
&-\tfrac{128 (2 z+1)}{(z^2-2 z-1)^3}+\tfrac{52}{3 (z+1)}-\tfrac{797929}{864 z}+\tfrac{12305}{72 z^2}+\tfrac{223}{6 z^3}+\tfrac{1334}{3 z^4}+\tfrac{1576}{3 z^5}) f_4+(-\tfrac{8 (127 z-206)}{z^2-2 z-1}\nonumber \\
&+\tfrac{128 (2 z+1)}{(z^2-2 z-1)^3}+\tfrac{16 (12 z+7)}{(z^2-2 z-1)^2}+\tfrac{6208}{9 (z-2)}-\tfrac{164}{3 (z-1)}+\tfrac{32}{3 (z-1)^2}-\tfrac{896}{9 (2 z-1)}+\tfrac{32}{3 (2 z-1)^2}-\tfrac{20}{z+1}\nonumber \\
&-\tfrac{16}{3 (z+1)^2}+\tfrac{1928}{3 z}-\tfrac{2000}{3 z^2}-\tfrac{168}{z^3}-\tfrac{2000}{z^4}) f_5+(\tfrac{-3075 z^2+14096 z-18121}{176 (z^3-4 z^2+3 z+4)}-\tfrac{2 (17 z^2-99 z-114)}{11 (z^3-5 z^2-8 z-4)}\nonumber \\
&-\tfrac{13}{3 (z-2)}-\tfrac{26}{3 (z+1)}+\tfrac{457}{16 z}-\tfrac{51}{4 z^2}+\tfrac{11}{z^3}+\tfrac{12}{z^4}) f_7+(\tfrac{20}{(z-2)^2}+\tfrac{96}{z-1}-\tfrac{128}{3 (z+1)}-\tfrac{8}{z}-\tfrac{68}{z^2}+\tfrac{48}{z^3}\nonumber \\
&-\tfrac{16}{z^4}+\tfrac{64}{z^5}-\tfrac{136}{3 (z-2)}) f_8+(\tfrac{4}{3 (z-2)^2}+\tfrac{1376}{3 (z+1)}+\tfrac{320}{3 (z+1)^2}-\tfrac{456}{z}+\tfrac{356}{z^2}-\tfrac{240}{z^3}+\tfrac{144}{z^4}-\tfrac{64}{z^5}\nonumber \\
&-\tfrac{8}{3 (z-2)}) f_9+(-\tfrac{2}{(z-2)^2}+\tfrac{128}{3 (z+1)}-\tfrac{46}{z}+\tfrac{38}{z^2}-\tfrac{32}{z^3}+\tfrac{24}{z^4}+\tfrac{10}{3 (z-2)}) f_{10}+(-\tfrac{6}{(z-2)^2}-\tfrac{2}{z}+\tfrac{2}{z^2}\nonumber \\
&+\tfrac{8}{z^4}+\tfrac{2}{z-2}) f_{11}+(-z^2+4 z-\tfrac{2}{z-2}-\tfrac{2}{z-1}+\tfrac{2}{z+1}-\tfrac{6}{z^2}+\tfrac{12}{z^3}-4+\tfrac{12}{z}) f_{12}+(-9 z-\tfrac{35}{2 (z-2)}\nonumber \\
&-\tfrac{18}{z-1}-\tfrac{2}{z+1}-\tfrac{9}{2 z}+\tfrac{73}{z^2}+\tfrac{10}{z^3}+\tfrac{76}{z^4}-8) f_{13}+(z^2+27 z+\tfrac{45}{2 (z-2)}-\tfrac{32}{z-1}+\tfrac{75}{2 z}+\tfrac{33}{z^2}-\tfrac{30}{z^3}\nonumber \\
&+\tfrac{36}{z^4}+8) f_{15}+(-\tfrac{22}{z+1}+\tfrac{22}{z}-\tfrac{32}{z^2}+\tfrac{36}{z^3}-\tfrac{16}{z^4}+2+\tfrac{4}{z-2}) f_{16}+(-z^2+2 z-\tfrac{10}{3 (z-2)}+\tfrac{64}{3 (z+1)}\nonumber \\
&+\tfrac{18}{z^2}-\tfrac{4}{z^3}-1-\tfrac{20}{z}) f_{18}+(-6 z^2-9 z-\tfrac{51}{z-2}+\tfrac{22}{z+1}-\tfrac{6}{z^2}-\tfrac{50}{z^3}+\tfrac{24}{z^4}-65+\tfrac{27}{z}) f_{19}\nonumber \\
&+(-\tfrac{5 z^2}{2}-\tfrac{89 z}{2}-\tfrac{115}{2 (z-2)}+\tfrac{55}{2 z}-\tfrac{20}{z^2}+\tfrac{14}{z^3}-\tfrac{8}{z^4}) f_{20}+(-\tfrac{64}{3 (z+1)}+\tfrac{22}{z}-\tfrac{36}{z^2}+\tfrac{40}{z^3}-\tfrac{16}{z^4}\nonumber \\
&-\tfrac{2}{3 (z-2)}) f_{21}+(-\tfrac{z^2}{2}-\tfrac{z}{2}+\tfrac{213}{2 (z-2)}-\tfrac{16}{z-1}-\tfrac{13}{2 z}-\tfrac{46}{z^2}+\tfrac{4}{z^3}-\tfrac{16}{z^4}+55) f_{22}+(\tfrac{2}{z-1}-\tfrac{10}{z}-\tfrac{16}{z^2}\nonumber \\
&-\tfrac{28}{z^3}-\tfrac{16}{z^4}+2+\tfrac{16}{z-2}) f_{23}+(4 z^2+20 z+\tfrac{48}{z-2}-\tfrac{24}{z^2}+36+\tfrac{24}{z}) f_{24}-\tfrac{16}{z-2}+\tfrac{24}{(z-2)^2}-\tfrac{278}{3 (z-1)}\nonumber \\
&-\tfrac{16}{2 z-1}-\tfrac{32}{(z^2-2 z-1)^2}-\tfrac{132}{z+1}+\tfrac{598}{3 z}+\tfrac{152}{z^2}+\tfrac{52}{3 z^3}+\tfrac{1432}{3 z^4}-\tfrac{352}{z^5}+\tfrac{384}{z^6}\Big\}\tfrac{64 \pi ^2}{3 \sqrt{3}}.
\end{align}
Here $z\equiv 2r-1$, and $f_i$ are defined as
\begin{align}
&f_1=i\pi,\quad f_2=\ln (2),\quad f_3=\ln (\tfrac{1-z}{2}),\quad f_4=\ln (\tfrac{1+z}{2}),\quad f_5=\ln (z),\nonumber \\
&f_6=\ln (\tfrac{z^2-3 z+4}{2}),\quad f_7=\sqrt{\tfrac{z^3-5 z^2-8 z-4}{(z-1) z^2}} \ln \Big(\tfrac{\sqrt{\tfrac{z^3-5 z^2-8 z-4}{(z-1) z^2}}-1}{\sqrt{\tfrac{z^3-5 z^2-8 z-4}{(z-1) z^2}}+1}\Big),\nonumber \\
&f_8=\tfrac{1}{2} (\tfrac{1-z}{2})^2 ((1-2 \ln (\tfrac{1-z}{2}))^2+\tfrac{\pi ^2}{3}+1),\quad f_9=\tfrac{1}{2} (\tfrac{1+z}{2})^2 ((1-2 \ln (\tfrac{z+1}{2}))^2+\tfrac{\pi ^2}{3}+1),\nonumber \\
&f_{10}=\tfrac{(1-z)^2}{2 (3-z) z}\ln (\tfrac{(1-z)^2}{1+z}) (\ln (\tfrac{16 (3-z)^2}{(1+z)^3 (1-z)^2 (2-z)^2})+4) +2 (\ln (\tfrac{z+1}{2})-2) \ln (\tfrac{z+1}{2})\nonumber \\
&\quad+\tfrac{1}{2 z}(\ln (\tfrac{(1+z) (1-z)^4}{16})-4)\ln (1+z) -2 \ln (\tfrac{(1-z)^2}{1+z}) \ln (\tfrac{3-z}{2-z}))\nonumber \\
&\quad+\tfrac{(1+z) (2-z) }{(3-z) z}({\rm Li}_2(-\tfrac{1}{2-z})-{\rm Li}_2(-\tfrac{(1-z)^2}{(1+z) (2-z)}))+\tfrac{\pi ^2}{6}+4,\nonumber \\
&f_{11}=-\tfrac{(6-7 z)}{2 z}\ln^2 (\tfrac{1-z}{2})+\tfrac{2 (2-3 z) }{z}\ln (\tfrac{1-z}{2})+\tfrac{(z-2)}{z}({\rm Li}_2(\tfrac{1}{2-z})-{\rm Li}_2(\tfrac{1-z}{2-z}))\nonumber \\
&\quad-\tfrac{(z-2) }{2 z}(\ln (2) (3 \ln (2)+4)-2 \ln (1-z) \ln (2 (2-z)))+\tfrac{\pi ^2}{6}+4,\nonumber \\
&f_{12}=C_0(0,\tfrac{1}{4} (z-1) z^2,-\tfrac{1}{4} (z-3) z^2,\tfrac{1}{4} (z-1)^2,\tfrac{1}{4} (z+1)^2,\tfrac{1}{4} (z+1)^2),\nonumber\\
&f_{13}=C_0(0,\tfrac{1}{4} (z-1)^2,\tfrac{1}{4} (-z^2+2 z+1),\tfrac{1}{4} (z+1)^2,\tfrac{1}{4} (z-1)^2,0),\nonumber\\
&f_{14}=C_0(0,\tfrac{1}{4} (z-1)^2,-\tfrac{1}{4} (z-1) (2 z^2-5 z+1),\tfrac{1}{4} (z+1)^2,\tfrac{1}{4} (z-1)^2,0),\nonumber\\
&f_{15}=C_0(0,\tfrac{1}{4} (z+1)^2,\tfrac{1}{4} (z-1) (z+1) (2 z-1),\tfrac{1}{4} (z-1)^2,\tfrac{1}{4} (z+1)^2,0),\nonumber\\
&f_{16}=C_0(-\tfrac{1}{4} (z-3) z^2,\tfrac{1}{4} (z-1)^2,\tfrac{1}{4} (-z^2+2 z+1),\tfrac{1}{4} (z+1)^2,\tfrac{1}{4} (z-1)^2,0),\nonumber\\
&f_{17}=C_0(-\tfrac{1}{4} (z-3) z^2,\tfrac{1}{4} (z-1)^2,-\tfrac{1}{4} (z-1) (2 z^2-5 z+1),\tfrac{1}{4} (z+1)^2,\tfrac{1}{4} (z-1)^2,0),\nonumber\\
&f_{18}=C_0(\tfrac{1}{4} (z-1) z^2,-\tfrac{1}{4} (z-3) z^2,0,\tfrac{1}{4} (z-1)^2,\tfrac{1}{4} (z-1)^2,\tfrac{1}{4} (z+1)^2),\nonumber\\
&f_{19}=C_0(\tfrac{1}{4} (z-1) z^2,\tfrac{1}{4} (z-1)^2,\tfrac{1}{4} (z-1)^2,0,0,\tfrac{1}{4} (z-1)^2),\nonumber\\
&f_{20}=C_0(\tfrac{1}{4} (z-1) z^2,\tfrac{1}{4} (z-1)^2,\tfrac{1}{4} (z-1) (z+1) (2 z-1),0,0,\tfrac{1}{4} (z-1)^2),\nonumber\\
&f_{21}=C_0(\tfrac{1}{4} (z-1) z^2,\tfrac{1}{4} (z-1)^2,\tfrac{1}{4} (z-1) (z+1) (2 z-1),\tfrac{1}{4} (z-1)^2,\tfrac{1}{4} (z-1)^2,0),\nonumber\\
&f_{22}=C_0(\tfrac{1}{4} (z-1) z^2,\tfrac{1}{4} (z+1)^2,\tfrac{1}{4} (-z^2+2 z+1),0,0,\tfrac{1}{4} (z+1)^2),\nonumber\\
&f_{23}=C_0(\tfrac{1}{4} (z-1) z^2,\tfrac{1}{4} (z+1)^2,\tfrac{1}{4} (-z^2+2 z+1),\tfrac{1}{4} (z+1)^2,\tfrac{1}{4} (z+1)^2,0),\nonumber\\
&f_{24}=C_0(\tfrac{1}{4} (z-1)^2,(z-1)^2,\tfrac{1}{4} (z-1) (z+1) (2 z-1),0,\tfrac{1}{4} (z-1)^2,\tfrac{1}{4} (z-1)^2),\nonumber\\
&f_{25}=C_0(\tfrac{1}{4} (z-1)^2,(z-1)^2,\tfrac{1}{4} (z-1) (z+1) (2 z-1),\tfrac{1}{4} (z-1)^2,0,0),\nonumber\\
&f_{26}=C_0((z-1)^2,\tfrac{1}{4} (z+1)^2,-\tfrac{1}{4} (z-1) (2 z^2-5 z+1),0,0,\tfrac{1}{4} (z+1)^2),
\end{align}
where $C_0$ is the standard Passarino-Veltman scalar three-point function.

\end{document}